\documentclass[a4paper,twoside,twocolumn,11pt]{article}
\usepackage[numbers]{natbib}
\usepackage{colordvi,color}
\usepackage{graphics,graphicx}
\usepackage{multirow}
\usepackage[font=small,format=plain,labelfont=bf,up,textfont={it,up,stretch=0.85}]{caption}

\usepackage{fancyhdr}
\usepackage{setspace}  
\usepackage{enumitem}

\usepackage{titlesec}
\titleformat{\section}{\normalfont\large\bfseries}{\thesection}{1em}{}
\titleformat{\subsection}{\normalfont\normalsize\mdseries}{\thesubsection}{1em}{}
\titleformat{\subsubsection}{\normalfont\normalsize\mdseries}{\thesubsubsection}{1em}{}
\titlespacing*{\section} {0pt}{3.5ex plus 1ex minus .2ex}{1.8ex plus .2ex}
\titlespacing*{\subsection}{0pt}{3.0ex plus 1ex minus .2ex}{1.ex plus .2ex}
\titlespacing*{\subsubsection}{0pt}{2.75ex plus 1ex minus .2ex}{0.75ex plus .2ex}
\titlespacing*{\paragraph}{0pt}{.5ex plus 1ex minus .2ex}{0.2ex plus .2ex}

\usepackage{mathptmx}

\def\msun{M$_{\odot}$}
\def\lya{\mbox{Ly$\alpha$}}
\def\asec{\ifmmode ^{\prime\prime}\else$^{\prime\prime}$\fi}
\def\amin{\ifmmode ^{\prime}\else$^{\prime}$\fi}
\def\degs{\ifmmode ^{\circ}\else$^{\circ}$\fi}
\def\fdg{\hbox{$.\!\!^\circ$}}          

\def\lsim{\mathrel{\rlap{\lower4pt\hbox{\hskip1pt$\sim$}}
    \raise1pt\hbox{$<$}}}                
\def\gsim{\mathrel{\rlap{\lower4pt\hbox{\hskip1pt$\sim$}}
    \raise1pt\hbox{$>$}}}                

\definecolor{lightblue}{rgb}{0.85,0.9,1}

\begin{document}

\cfoot[\fancyplain{}{}]{\fancyplain{}{}}

\vspace{-0.5cm}
\parindent=0pt

\begin{figure}[bh]
\vspace{-2.9cm}
\hspace*{-2.4cm}\includegraphics[width=1.25\textwidth]{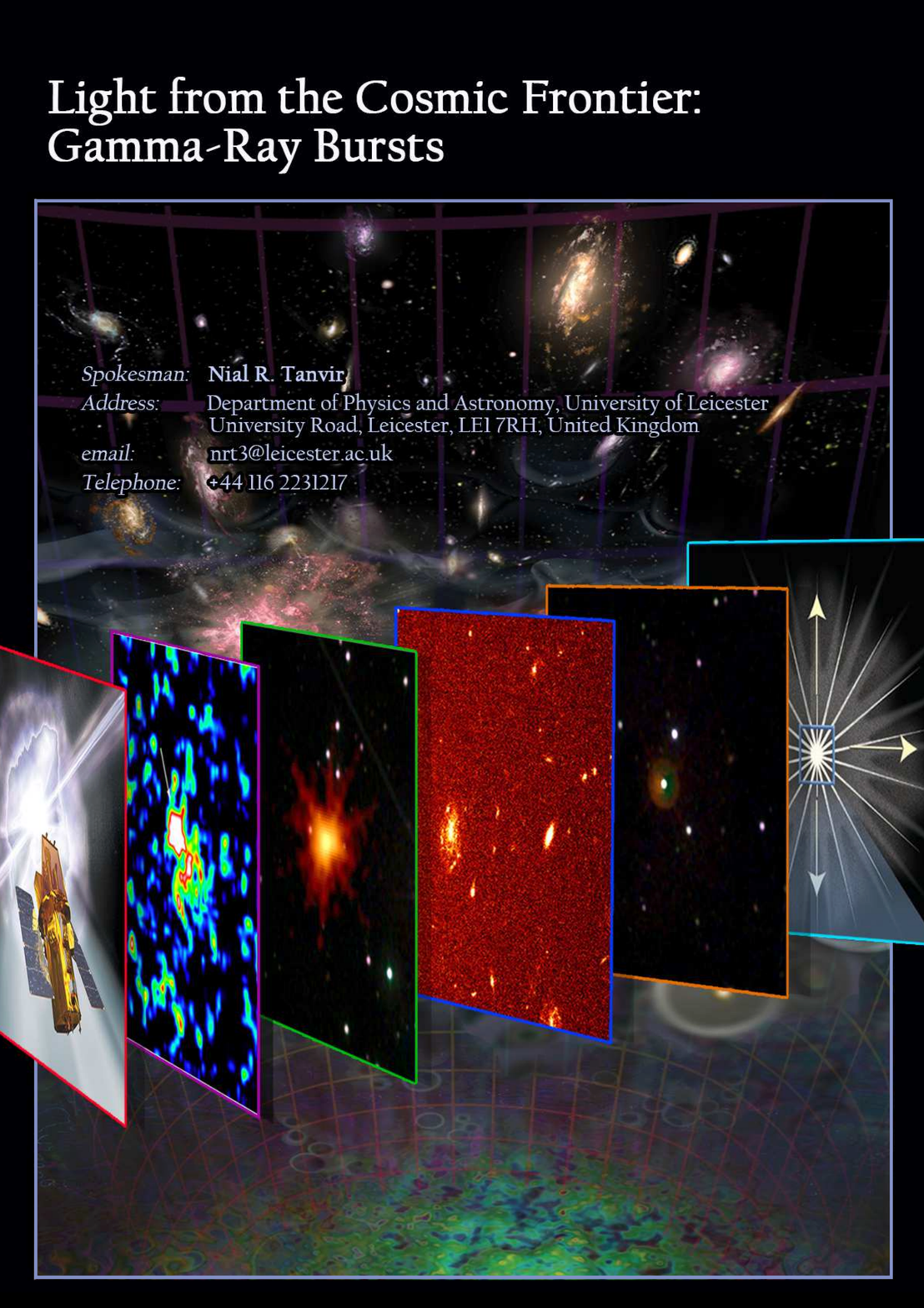}
\end{figure}

\clearpage

\renewcommand{\baselinestretch}{1.1}\normalsize

\parbox{\textwidth}{
\medskip
{\Large Authors:}

\medskip

\begin{flushleft}
{\large
L. Amati (INAF/IASF Bologna),
~~J.-L. Atteia (IRAP/CNRS/UPS Toulouse),\linebreak
L. Balazs (Konkoly Obs.), 
~~S. Basa (LAM Marseille),
~~J. Becker Tjus (Univ. Bochum),\linebreak
D.F. Bersier (Liverpool Univ.), 
~M. Bo\"er (CNRS-ARTEMIS Nice),
S. Campana (INAF/OAB Brera),\linebreak
B. Ciardi (MPA Garching), 
~~S. Covino (INAF/OAB Brera),
~~F. Daigne (IAP Paris), \linebreak
M. Feroci (INAF/IAPS Rome), 
~~A. Ferrara (SNS Pisa),
~~F. Frontera (INFN/Univ. Ferrara), \linebreak
J.P.U. Fynbo (DARK Copenhagen), 
~~G. Ghirlanda (INAF/OAB Brera), \linebreak
G. Ghisellini (INAF/OAB Brera),
~~S. Glover (Uni. Heidelberg), 
~~J. Greiner (MPE Garching),\linebreak
D. G\"otz (CEA Saclay),
~~L. Hanlon (UCD Dublin),
~~J. Hjorth (DARK Copenhagen), \linebreak
~~R. Hudec (AI AS CR Ond\v{r}ejov \& CTU Prague),
~~U. Katz (Univ. Erlangen),\linebreak
S. Khochfar (Univ. Edinburgh), 
~~R. Klessen (Univ. Heidelberg),
~~M. Kowalski (Univ. Bonn), \linebreak
A.J. Levan (Univ. Warwick), 
~~S. McBreen (UCD Dublin), 
~~A. Mesinger (SNS Pisa), \linebreak
R. Mochkovitch (IAP Paris), 
~~P. O'Brien (Univ. Leicester), 
~~J.P. Osborne (Univ. Leicester),\linebreak
P. Petitjean (IAP Paris), 
~~O. Reimer (Univ. Innsbruck),
~~E. Resconi (TU M\"unchen),\linebreak
S. Rosswog (Univ. Stockholm), 
~~F. Ryde (KTH Stockholm),
~~R. Salvaterra (INAF/IASF Milano),\linebreak
S. Savaglio (MPE Garching), 
~~R. Schneider (INAF/Oss. Rome),
~~G. Tagliaferri (INAF/OAB Brera),\linebreak
A. van der Horst (Univ. Amsterdam)
}
\end{flushleft}
}

\vspace{3.cm}

\parbox{\textwidth}{
{\Large Supporters:}

\medskip

\begin{flushleft}
{\large
M. Ackermann  (DESY Zeuthen),
~~Z. Bagoly (E\"otv\"os Univ.),
~~E. Bernardini (DESY Zeuthen),\linebreak 
J.H. Black (Chalmers Univ. of Techn.),
~~P. Clark (Uni. Heidelberg),
B. Cordier (CEA Saclay),\linebreak
~~J.-G. Cuby (LAM Marseille), 
~~F. Ferrini (Univ. Pisa),
C. Finley (Stockholm Univ.),\linebreak 
~~S. Klose (Tautenburg Obs.), 
~~A. Klotz (IRAP/CNRS/UPS Toulouse),\linebreak
T. Kr\"uhler (DARK Copenhagen),
~~N. Langer (Univ. Bonn),
~~K. Mannheim (W\"urzburg Univ.), \linebreak
E. Nakar (Wise Obs.),
~~C.-N. Man (ARTEMIS, CNRS/OCA Nice),\linebreak
M. Pohl (DESY Zeuthen, Potsdam Univ.),
~~P. Schady (MPE Garching),
~~S. Schanne (CEA Saclay),\linebreak
V. Springel (ITS Heidelberg), 
~~P. Sutton (Cardiff Univ.),
~~N. van Eijndhoven (Brussel Univ.), \linebreak
J.-Y. Vinet (ARTEMIS, CNRS/OCA Nice), 
~~A. Vlasis (CPA Leuven),\linebreak
D. Watson (DARK Copenhagen),
~~K. Wiersema (Univ. Leicester)
}
\end{flushleft}

\vspace{2.cm}

{\Large Supporters from non-ESA member states:}

\medskip
{\large
V. Bromm (Univ. Texas Austin), ~~N. Gehrels (GSFC),
~~N. Kawai (Tokyo Inst. Techn.)
}

}

\clearpage

\renewcommand{\baselinestretch}{0.85}\normalsize

\cfoot[\fancyplain{}{\thepage}]{\fancyplain{}{\thepage}}
\setcounter{page}{1} 

\section{Executive Summary}

Gamma-Ray Bursts (GRBs) are the most powerful cosmic explosions since the 
Big Bang,  and thus act as signposts throughout the distant Universe.
Over  the last 2 decades,
these ultra-luminous cosmological  explosions have been transformed from a
mere curiosity  to essential tools for the study of high-redshift
stars and galaxies, early structure formation and the evolution of
chemical elements.
In the future, 
GRBs will likely provide a powerful probe of the 
epoch of re-ionisation of the Universe, constrain the properties of the 
first generation of stars (Fig.~1), and play an important role in the
revolution of multi-messenger astronomy by associating 
neutrinos or gravitational wave (GW) signals with GRBs.

Here, we describe the next steps needed to advance the GRB field,
as well as the potential of GRBs for studying the Early Universe
and their role in the upcoming multi-messenger revolution.
We identify the following fundamental questions as the prime science
drivers for the next two decades:
\begin{itemize}[leftmargin=*]
\vspace{-0.22cm}\item When did the first stars form, what are their properties,
  and how do Pop III
  stars differ from later star formation in the presence of metals?
\vspace{-0.22cm}\item When and how fast was the Universe enriched with metals?
\vspace{-0.22cm}\item How were the first structures formed which then 
 developed into the first galaxies?
\vspace{-0.22cm}\item How did reionisation proceed as a function of environment,
and was radiation from massive stars its primary driver?
\vspace{-0.22cm}\item What is the relation between GRB rate and star 
  formation rate, and what is its evolution with time?
What is the true redshift distribution and 
corresponding   luminosity function of long GRBs?
\vspace{-0.22cm}\item How are $\gamma$-ray and neutrino
  flux in GRBs related, and how do neutrinos from long GRBs constrain the 
  progenitor and core-collapse models?
\vspace{-0.22cm}\item Can short-duration GRBs be unambiguously linked to 
  gravitational wave signals, and what do they tell us about the neutron 
  star merger scenario?
\vspace{-0.22cm}\item What are the electromagnetic counterparts to 
  gravitational waves and neutrino bursts?
\vspace{-0.2cm}
\end{itemize}

These questions relate directly to the Cosmic Vision theme \#4,
``{\em How did the Universe originate and what is it made of?}'',
in particular to 3 out of the 8 goals:
{\bf (1)} 
  {\em Find the first gravitationally-bound structures that were assembled 
  in the Universe -- precursors to today's galaxies and cluster 
  of galaxies -- and trace their evolution to the present.}
  Since GRBs can be detected from extreme
  distances ($z\sim$30--60 \cite{nab07}), accurate
  localisations provide the best-possible pointers
  to the first stars, and to the proto-galaxies where they form. 
  This will find the first black holes (BH), likely to be the seeds
  of the super\-massive BHs which dominate the X-ray luminosity
  of the current Universe.

\begin{figure}[th]
\vspace{-0.2cm}
\includegraphics[width=1.0\columnwidth]{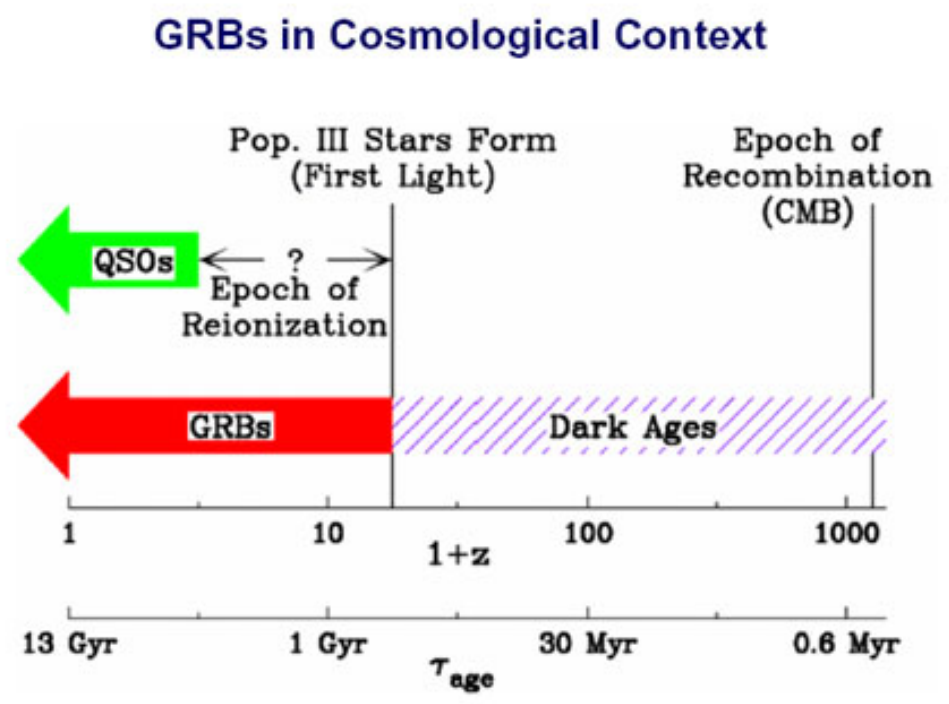}
\vspace{-0.8cm}
\caption{\colorbox{white}{GRBs in the cosmological context. (From \cite{LaR2000})}}
\label{highzGRBs}
\end{figure}

{\bf (2)} 
   {\em ...Trace the life cycle of chemical elements through cosmic history.}
Since long GRBs are produced by massive stars \cite{hjo03},
    which have short life-times, the GRB rate traces the 
    star formation (SF) rate.  The  extraordinary
  brightness and minimised absorption bias make GRBs particularly
  useful SF indicators. Optical/NIR afterglow spectroscopy 
  allows to measure line-of-sight metallicity at 
  exquisite detail, and map the cosmic chemical evolution
  with high-$z$\, GRBs. Short GRBs are likely linked to the formation of
  the heaviest elements in the Universe, such as platinum.

{\bf (3)} 
  {\em ...Examine the accretion processes of matter falling into black holes...,
   and look for clues to the processes at work in gamma-ray bursts. }
  The prompt $\gamma$/X-ray emission of GRBs, combined with
  polarisation measurements provides direct clues on the accretion
  and jet ejection processes. Optical/NIR absorption line diagnostics
  will link this to progenitor properties, thus allowing us to understand
  the basic picture of GRB-production.

To fully utilise GRBs as probes of the
early Universe and/or as multi-messengers requires 
(i) a detection system for 5000 GRBs, among them 50 GRBs at $z>10$,
(ii) a means to localise them quickly and accurately, (iii) 
instrumentation to determine their redshift quickly, preferentially
on the fly, and (iv) low-latency communication to the ground.

Instead of proposing a single strawman mission concept, we describe
the instrumentation needed to answer each of the above fundamental 
questions, and describe options how to realise the measurements
depending on the priorities among those questions.

The authors of this WP agree in the choice of emphasising
the role GRBs can and will play in the study of the high-redshift Universe,
but also recognise the huge potential of GRBs for
the coincident detection of GW or neutrinos
and electromagnetic signals \cite{bbm13}.
It may confirm the basic model of the short GRBs, 
finally clarify the origin of the heaviest elements, and allow
for a precise measurement of the expansion rate of the Universe.

\clearpage

\section{The astrophysics landscape until 2030}

The landscape of astronomy 15-20 years from now will be 
dominated by a wealth of new facilities across the
entire electromagnetic (EM) spectrum and beyond, as non-photonic sources 
open a previously inaccessible window into many of
the most extreme regions of the Universe. It is against this landscape 
that new space missions and ground-based
facilities must be measured and judged.
While any attempt to prescribe precisely the likely scientific frontiers at 
this time is fraught with uncertainty, a variety of
possibilities bear consideration. We outline a range of facilities that may 
 be operational in this time scale, along with
their contribution to the science questions.
A GRB-focussed mission would provide a huge step in our understanding
of the early Universe, impossible by any of the facilities in the planning,
and at the same time would enable some of these planned facilities
to perform science that would be otherwise impossible.

GRBs are the most luminous sources on the sky, 
releasing in less than a minute the energy output of the Sun over 
its entire life.  Several GRBs occur each day, and thus GRBs act as 
frequently available signposts throughout the Universe. 
Two sub-groups of GRBs are distinguished according to their duration (Fig.~2):
(i) Long-duration GRBs ($>$2 s) are firmly linked to the collapse 
of massive stars,
thus probing sites of star formation with little delay, as the star's lifetimes
are measured in megayears and not gigayears. GRBs have been seen
up to the highest measured redshifts.
(ii) Short-duration GRBs likely originate
from the merging of compact stars and are expected to produce strong 
gravitational waves.
Both types of GRBs are powerful neutrino sources. As stellar sized
objects at cosmological scales, they connect different branches of 
research and thus have a broad impact on present-day astrophysics.

\begin{figure}[th]
\includegraphics[width=0.99\columnwidth]{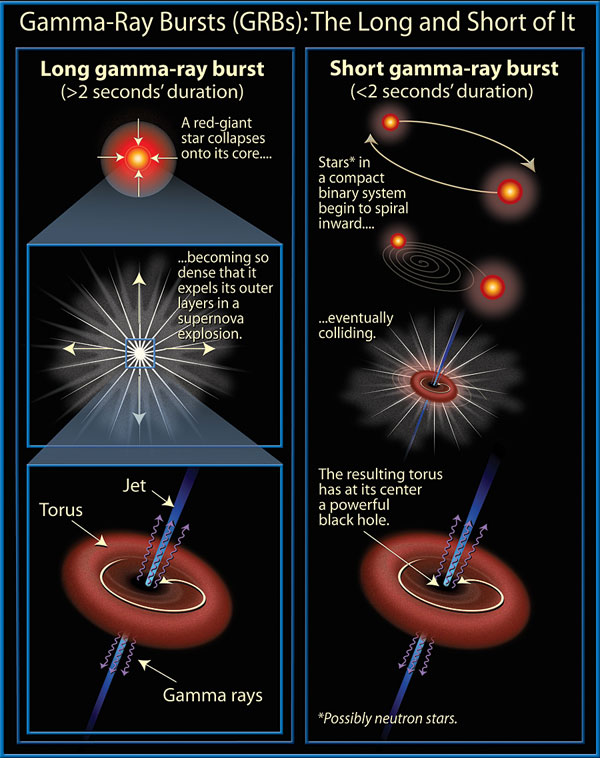} 
\caption{Long-duration GRBs (left) are thought to originate in the 
collapse of massive stars, while short-duration GRBs are likely
produced in the merger of two compact objects. Both scenarios result
in a relativistic jet which is responsible for producing the $\gamma$-ray
emission.  (From \cite{slgrbs})}
\label{GRBdur}
\end{figure}

\subsection{High-Energy satellites}

Over the next few years, progress on GRBs is likely to remain driven 
by the {\em Swift} mission. Its launch heralded an 
unprecedented period of progress towards GRB progenitors, as well as 
highlighting the varied and diverse high-energy sky in ways that were 
unanticipated prior to its launch. {\em Swift} achieved this due to
a combination of a broad compliment of instruments dedicated to 
GRB detection and follow-up, and the implementation of a novel autonomous
rapidly-slewing spacecraft. It has found the first
GRBs at $z>6,7,8\,{\rm and}\, 9$, pinpointed the 
locations of short-GRB afterglows, identified nearby GRBs with and 
(importantly) without supernovae. Despite its 8 yr in orbit, it continues
to discover new populations of high-energy transients in previously 
unexplored parameter space. 
{\em Swift} was joined in 2008 by the {\em Fermi} Gamma-ray Telescope 
-- a powerful satellite 
with an unparalleled  spectral range, opening new
insights into the nature of the $\gamma$-ray emission from GRBs, and enabling 
sensitive tests of differing models for quantum gravity. 
Real-time GRB detections are also provided by
{\em INTEGRAL}, {\em AGILE}, {\em Suzaku}, {\em MAXI/ISS} and 
the interplanetary network (IPN) satellites.

These missions are all working well at present, but have finite lifetimes, 
governed both by orbital decay, instrument lifetime and, perhaps more 
importantly, financial constraints. It is unlikely that any of them will 
still be in 
operation well into the 2020s. Our window onto the transient high-energy sky 
thus revolves around new initiatives.
Those likely in the interim period until 2028 are specialised 
instruments, often with 
lower sensitivity than {\em Swift} which will focus on individual science 
questions.

Four larger scale missions are the approved Indian {\em Astrosat}, the
Japanese {\em Astro-H}, and the German/Russian {\em SRG}, 
as well as the planned  French/Chinese {\em SVOM}.
{\em Astrosat} is a multi-wavelength observatory covering the UV to hard X-ray
bands scheduled for launch within a year, and may be expected to detect 
of order a dozen GRBs per year with its Scanning Sky Monitor. 
{\em Astro-H} is scheduled for launch in 2015, and might be able to obtain
high-resolution spectra of GRB afterglows if target-of-opportunity
observations can be rapidly scheduled.
{\em SRG} will perform a sensitive all-sky survey in the 0.3--12 keV
band with the {\em eROSITA} and ART-XC telescopes, starting
in 2015.
In particular, {\em eROSITA} \cite{predehl2010} with its good sensitivity is 
expected to detect 4--8 GRB afterglows per year \cite{kss12}, over a 
mission lifetime of at least 4 years.
{\em SVOM} in many ways is modelled on the remarkable success 
of {\em Swift}, carrying $\gamma$-ray, X-ray and optical telescopes.
The softer  response (triggering at lower energy) 
and larger, red optimised optical 
telescope may enhance the recovery fraction for high-$z$ GRBs.

Other future high-energy missions remain at an earlier stage of development, 
although there are plans in progress to launch small to moderate size 
detectors either as stand-alone missions or via the ISS.
{\em LOFT} is currently under 
consideration for ESA's M3 launch slot as primarily a timing experiment, 
providing spectral sensitivity and timing resolution much better than 
{\em RXTE}.  Its wide-field monitors make it a capable GRB detector. 
However, 
it has no automated slewing capability.
In a 4 year mission it will see only $<2$ GRBs at $z>8$, even under 
optimistic assumptions \cite{Amati2013}. 
Japan is planning an upgrade of {\em MAXI} on the ISS within the next few years,
and NASA has recently accepted {\em NICER} \cite{gao12}, an X-ray timing and 
spectroscopy experiment for the ISS, with a launch date in 2017
which potentially could be used for GRB afterglow observations.
The Russian space agency is planning for a small GRB mission within the
next year with {\em UFFO}-pathfinder, a rapid detection system for prompt 
optical emission which later might evolve into a larger {\em UFFO} mission
\cite{pbb13}.
Planned for launch in the next years on the Chinese Tiangong-2 is
a hard X-ray polarimeter for the study of GRBs, though this
relies on the localisation and spectral measurements by a different
satellite.

\subsection{Multi-wavelength \& multi-messenger domain}

Outside of the high energy arena, the next years should see the long awaited 
start of routine multi-messenger astronomy, for which
neutrinos from SN~1987A offered the first hints. The power of this 
non-photonic messenger, and gravitational waves as well, is to probe into 
highly enshrouded environments, invisible to electromagnetic observers.
Both messengers are currently the subject
of major investments, still have to reach a positive detection of signals 
from GRBs,
but are expected to remedy this situation in the next decade .

Firstly, the upgrades to the {\em LIGO} and {\em VIRGO} interferometers will 
reach the point at which routine astrophysical detections of
gravitational waves become reality \cite{Siellez2013}. 
This point should be reached towards the end of this decade \cite{Aasi2013}.
Further away is a next generation of GW interferometer known as the
Einstein Telescope (ET) \cite{Sathy12} 
with a target operational date in the mid-2020's. 
Since these detectors measure gravitational
wave strain, the observational horizon scales linearly with the sensitivity
(unlike the inverse square law for electromagnetic detectors).
ET will be capable of seeing compact binary mergers
to $z \sim 3$ (compared to the $z \sim 0.1$ for the next generation detectors).
It will provide detection rates of 10$^4$ (10$^5$) for binary BH (NS)
mergers \cite{aba10}, enabling
detailed population and evolution studies. 
Mergers also provide a precise gravitational wave
luminosity distance, giving a powerful probe of cosmology that can
independently measure
$H_0$, $\Omega_M$, $\Omega_\Lambda$, $w$ and $\dot{w}$. 
However, positional accuracy will be poor, even for ET operating in 
conjunction with further upgraded {\em ALIGO/AVIRGO} detectors.
An EM trigger-system will therefore be needed to pinpoint source locations 
to an accuracy that allows
measuring their redshifts, since it is the comparison of the redshift to the
GW-determined distance that enables cosmological studies.

Secondly, IceCube, the largest neutrino telescope built so far, 
has been in operation since two years.
First studies now reach beyond the level of
predicted neutrino fluxes, but yielded no detection of GRB neutrinos
so far \cite{Abbasi2012}. 
The recently announced first hint of an astrophysical signal seen by
IceCube provides great prospects for the identification of cosmic ray sources
\cite{Whitehorn2013}.
Possible reasons could be the choice of parameters
for the standard neutrino flux calculations, in particular the Lorentz
factor of the source, and the relation between accelerated electrons and
protons. A more detailed treatment of the microphysics leads to a
reduction of the general flux at fixed parameters, but does not take into
account the general assumption that GRBs are the sources of ultra-high
energy cosmic rays (UHECRs) \cite{HBW12}. 
Further studies
will have to show if there is any significant spatial or temporal
clustering that could be connected to GRBs and will help to
study GRBs as possible cosmic ray sources.
The future European neutrino telescope KM3NeT, to be deployed within this
decade in the Mediterranean Sea, should provide another sensitivity boost,
so that expectations remain high. 

\begin{figure}[ht]
\vspace{-0.7cm}\hspace{-0.5cm}
\includegraphics[width=1.27\columnwidth, angle=0]{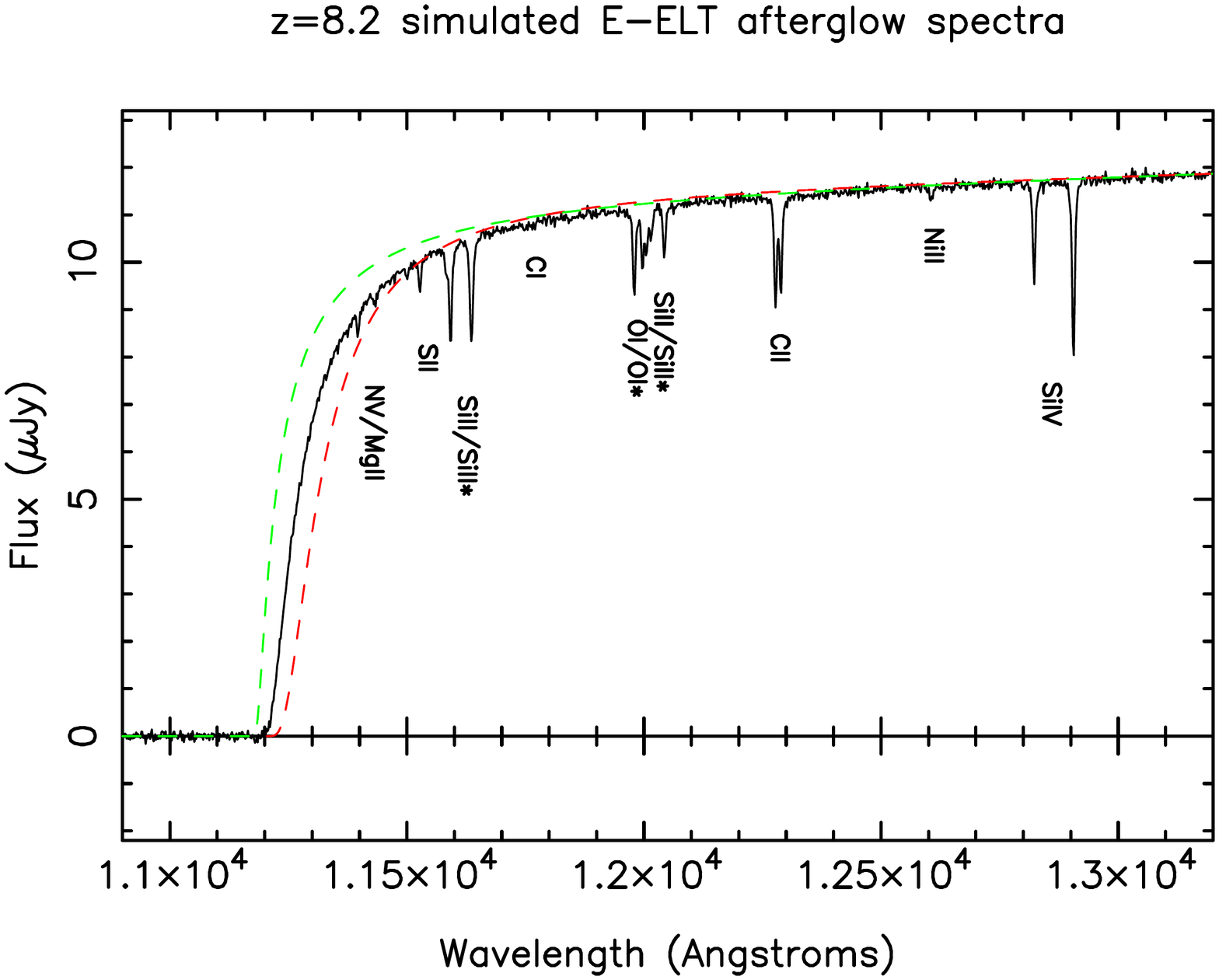}
\vspace{-0.9cm}
\caption{Simulated spectrum (solid line) around the Ly-$\alpha$ break
showing the quality of data which would be obtained with a 30--40\,m
telescope such as the proposed E-ELT for an afterglow with magnitude
approximately the same as that obtained for GRB 090423 (as observed 
by the VLT) 
The host galaxy was chosen to have an HI column density of 
10$^{21}$ cm$^{-2}$ and a
metallicity of 1/10 $Z_\odot$, and the IGM to be 100\% neutral.
The green dashed line shows a model with just a neutral IGM (with
redshift fixed at that given by the host metal lines).
High S/N data can be used to decompose
IGM and host galaxy contributions (red dashed line), thus determining
each with good precision.  This simulation also shows the excellent
measurements of metal abundances that could be achieved with such
observations.}
\label{EELTspec}
\end{figure}

The end of this decade (or the start of the next) will see the advent of 
the James Webb Space Telescope (JWST) and
large ground-based optical telescopes (ELT's).
These are observatories rather
than dedicated missions, with a science remit from exo-planets to cosmology. 
Central to the science case for each is the study
of the early Universe. These large telescopes will be used to pin-point some 
of the most distant galaxies yet observed as well as providing
spectroscopic capability beyond the limit of HST photometry. 
Nonetheless, even with these next
generation facilities, spectroscopic studies remain challenging. 
If the faint end 
slope of the galaxy luminosity function is genuinely very steep 
\cite{Bouwens2012, Tanvir2012}
then even these facilities will not probe it far down.
If the first stars form in relatively faint and low mass haloes, it is quite 
likely that they will not be found directly by either facility, even in
their deepest fields \cite{john2008}. 
As we will describe later, GRBs offer a route around this problem.
The 30--40\,m telescopes will be able to provide unique information
on the chemical enrichment and re-ionization history {\em if} they can
be fed with accurate locations of high-$z$ GRB afterglows (see 
Fig.~\ref{EELTspec} for a simulated E-ELT/HARMONI spectrum).

This period will also mark the launch of ESA's GAIA and Euclid satellites. 
GAIA is primarily an astrometry mission, but due to the temporal sequence of
its sky scans it will detect a large number of transients, among those
up to 40 GRB afterglows \cite{rpd08,souz12}.
Euclid is
predominantly aimed at providing precision measurements of cosmological 
parameters via weak lensing and baryon acoustic oscillations. The deep survey
should reach optical/NIR magnitudes of $VYJH \sim 24$ over half of the 
sky by 2027,
and may turn up a reasonable fraction ($\sim 30\%$) of low-redshift GRB hosts.

The Large Synoptic Survey  Telescope (LSST) is due to start operation 
around 2022. It will locate of order 10$^6$ transients per night.
High-energy coverage of a good fraction
of these transients would be of significant interest to the community
providing distinction between orphan GRB afterglows, tidal disruption flares, 
extreme supernovae, radioactively powered transients from
GW sources and other yet un-imagined transients. 
While LSST is expected to discover 4 GRB afterglows per night \cite{rpd08},
it will be limited to $z<7$ due to its filter set.

There are also significant ground-based investments across the
electromagnetic spectrum, from the high frequency of the 
Cherenkov Telescope Array (CTA) to the low frequency radio arrays associated 
with the Square Kilometer Array (SKA). CTA will be sensitive to the highest
energy $\gamma$-rays ($>10$ TeV in some cases), and will probe high-energy 
emission from GRBs and their shocks in
the first minutes after the bursts. The properties of the bursts at such 
high energy remain poorly understood at present, although
the recent GRB 130427A was detected up to 120 GeV (rest-frame; \cite{zrk13}).
Assuming the spectral-temporal extrapolations from presently
detected GRBs by Fermi-LAT, CTA might detect just a few GRB/yr \cite{gbc13}, 
but its orders-of-magnitude better sensitivity on short timescales 
compared to  
Fermi-LAT in the overlapping energy regime will provide a vast amount of
photon data allowing to sensitively probe spectral-temporal evolution of 
GRBs at the upper end of the accessible electromagnetic spectrum.
This will provide a handle on the prompt emission properties,
and will be a powerful complement to our proposed mission -- importantly, 
CTA will have a narrow field of view, and so will require triggers 
in order to re-point at GRBs.
CTA should be operational towards the end of the decade. We may
gain a somewhat earlier insight of the high-energy properties of GRBs via HAWC 
(the High Altitude Water Cherenkov Gamma-Ray observatory), which
is already partially operational, and has
the ability to trigger on $>1$ TeV $\gamma$-ray photons across 15\% of the sky,
(though it lacks the sensitivity of CTA and its
effective area decreases rapidly away from zenith).

Moving to longer wavelengths, the Atacama Large Millimeter Array (ALMA) 
will continue to be a workhorse instrument
for astronomy, and its unique sensitivity to warm and cool dust in distance 
galaxies  provides a means of probing the 
nature of the earliest galaxies, in particular of the highest redshift GRB 
hosts detected. In addition, ALMA will allow us to map systematically the
GRB afterglow emission near its spectral peak, thus providing 
beaming-independent energy estimates. 
Finally, SKA, which should be operational with a partial array around 2020, 
will provide 
new insights into the formation of the first structures in the Universe
and the re-ionisation through mapping the 21 cm line emission at 
different epochs. In the GRB field, SKA as well as its predecessors LOFAR,
MWA and PAPER will be powerful facilities to study
the radio afterglow emission, and will provide unique
insights into the physics and environment properties of these sources. 
We expect that SKA will be sensitive enough to detect all afterglows of
GRBs of a next generation $\gamma-$ray detector, and
for 50\% of those, 
will allow the estimate of their true (collimation corrected) 
energetics through late time ($>$100 days) radio calorimetry \cite{Ghirlanda13}.

\clearpage

\section{Open Questions}

\subsection{GRBs and the Early Universe}

\subsubsection{GRB observability}

The identification of very high redshift ($z>7$) sources is challenging due 
to their
great luminosity distances, and the difficulties of observing in the near-IR 
(NIR) from the ground.
This is exacerbated by the effects of hierarchical structure growth, which means
that galaxies start out increasingly small and faint intrinsically, and 
that bright quasars are exceptionally rare at $z>7$.
The scientific importance of studying this era has motivated very large
investments (or planned investments) in new NIR facilities
(e.g. aboard JWST) that are expected to detect sources at redshifts 
up to $z\sim13$ ($H$-band dropouts), 
but even they will struggle to find, much less confirm, galaxies beyond this.

To study the origin of the first stars and luminous structures in the universe,
observational advances to redshifts exceeding $z\sim13$ are essential.
The exploration of this challenging high-$z$ realm may be enabled by sources 
that are very bright, and have emission predominantly in the high energy 
regime, namely GRBs. 
Their specific advantages are:
(i) they likely exist out to the highest redshifts due to their creation
  in the deaths of massive stars,
(ii) the brightest bursts can easily be detected at the highest redshifts 
  due to their huge intrinsic luminosities and energy spectra peaking at
$\sim$100--300~keV;
(iii) their pan-chromatic afterglows can also be extraordinarily bright, providing 
backlights for detailed spectroscopy which is otherwise unprecedented at 
such distances; 
(iv)  they probe the epoch we seek to understand as their 
progenitor stars are likely representative of those responsible for the 
reionisation of the 
Universe: their current distance record is $z\approx9.4$ \cite{clf11}; and
(v) a favourable relativistic
k-correction implies that they do not get fainter beyond $z$$\sim$3.
Yet, present and near-future ground- and space-based capabilities are limiting
the measurement of redshifts at $z$$\sim$13 (as $H$-band drop-outs),
and their afterglows above 2.5 $\mu$m 
are too faint by many magnitudes for 8--10\,m telescopes.

To fully utilise GRBs as probes of the
early Universe one must localise large samples quickly and accurately, and 
be able to identify which of these are worth the valuable 30--40\,m telescope
time for detailed study, implying the determination of their (at least 
photometric) redshifts onboard.

\subsubsection{Structure formation scenarios}

From studying the cosmic microwave background we know that the Universe 
started out very simple. It was by and large homogeneous and isotropic, 
with small fluctuations that can be described by linear perturbation analysis. 
The present Universe, on the contrary, is highly structured and 
complicated. Cosmic evolution is thus a progression from simplicity to 
complexity, with the formation of the first stars and protogalaxies marking 
a primary milestone in this transition. Compiling and characterising a 
sample of very high redshift GRBs will help us directly probe this key 
phase of cosmic structure formation, as follows.

The first stars that give birth to high-$z$ GRBs must form out of gas 
that collected inside dense dark matter (DM) potential wells. 
Structure formation in a cold dark 
matter (CDM)-dominated Universe is "bottom up," with low-mass halos 
collapsing first. In the current concordance cosmology, with densities in 
CDM and dark energy of ($\Omega_{\rm M}$,$\Omega_{\Lambda}$) $\approx$ (0.3,0.7) 
as emerged from WMAP and Planck, DM halos 
with masses of 10$^5$--10$^6$ \msun\ \cite{teg97, ybh04} form 
from $\sim$3 $\sigma$ peaks of the initial primordial density field 
as early as $z \sim 25$. It is natural to identify these condensations 
as the sites where the first astrophysical objects, including the first 
massive stars, were born. Thus, one expects to find 
GRBs out to this limiting redshift but not beyond.

While the standard CDM model has been remarkably successful in
explaining the large-scale structures in the Universe and the
cosmic microwave background, some discrepancies remain at small scales,
$\lsim 1$ Mpc. Proposed alternatives are either baryonic feedback
or Warm DM (WDM; $\sim$ keV particles) models \cite{kyk13}.
In the latter case, the resulting effective pressure and free-streaming
would decrease structures on small scales \cite{bho01}. If indeed DM was 'warm',
the high-redshift Universe would be rather empty, such that even a 
single GRB at $z > 10$ would already provide strong constraints
on the WDM models \cite{mes05}.
Present constraints rule out WDM particles with masses smaller than
1.6--1.8 keV at 95\% confidence level, but depend on assumptions 
on the slope of the luminosity function and the GRB to SFR rate ratio.
Any improvements on these constraints requires
a substantially larger number of GRBs with measured redshifts at 
$z \gsim 5$ \cite{smf13}.

On a similar note, GRBs might be used to get independent constraints on
the amount of primordial non-Gaussianity in the density field \cite{Maio12}.
Deviations from the Gaussian case can only be found at high $z$.

Measurements of a statistically significant sample of GRBs 
(minimum $\sim$50)  at $z>10$ will therefore 
help to answer the question: 

\colorbox{lightblue}{\parbox{0.98\columnwidth}{
\textcolor{blue}{\bf How were the first structures 
formed which then developed into the first galaxies?}}}

\subsubsection{When and how did the first stars form?}

The nature of the first stars  in the Universe, and understanding how their  
radiative, chemical and mechanical feedback drove subsequent galaxy 
evolution, provide one of the grand challenges of
modern cosmology \cite{byh09}.
The earliest generations of stars ended the so-called cosmic dark ages and 
played a key role in the metal enrichment and reionisation of the Universe, 
thereby shaping the galaxies we see today \cite{byh09, bl04, glo05, CiF05}.
These so-called 
Population III (or Pop III) stars build up from truly metal-free primordial gas 
at extremely high redshift. They have long been thought to live 
short, solitary lives, with only one extremely massive star with about 
100 solar masses or more forming in each DM halo 
\cite{abn02, bcl02, yoh06, osn07}.
However, the most recent calculations \cite{cgs11a, gbc12, sgk13}
suggest that 
Pop III stars formed as members of multiple stellar systems with separations 
as small as the distance between the Earth and the Sun 
\cite{tao09, sgb10}.
Although these recent fragmentation calculations suggest an initial mass 
function (IMF) that reaches down to sub-solar values, most of the material 
is probably converted into intermediate mass stars with several tens of 
solar masses \cite{cgs11b, sgs11}.
This agrees with 
the analysis of abundance patterns of extremely metal-poor stars in the 
Galactic halo \cite{BeC05},
which requires a minimum level of enrichment to form low-mass and 
long-lived stars \cite{ssf07} and is consistent with 
enrichment from core collapse supernovae of stars in the intermediate mass 
range $20 - 40\,$M$_{\odot}$ rather than from pair-instability supernovae 
of very massive progenitors with $\sim 200\,$M$_{\odot}$ 
\cite{Tum07, tun07, iut09, HeW10, jab10}.

Second generation stars, sometimes termed Pop II.5 stars, have formed from 
material that has been enriched from the debris of the first stars. Unlike 
the very first stars, for which we have no direct detections yet, low-mass 
members of the second generation may have already been found in surveys looking 
for extremely metal-poor stars in our Milky Way and neighbouring satellite 
galaxies. The relative fraction of high-mass stars amongst Pop II.5 stars 
is still unknown. It is a key question in early galaxy formation to 
understand the transition from truly primordial star formation to the 
mode of stellar birth we observe today \cite{sfs03, sol12}. 
When and where did this transition 
occur? Was it smooth and gradual or rather sudden and rapid? It is therefore 
important to learn more about the IMF of the first and second generation 
of stars and to find observational constraints on the star formation process 
at different redshifts. This would culminate in the more general question:
\colorbox{lightblue}{\parbox{0.98\columnwidth}{
\textcolor{blue}{\bf When did the first stars form, what are their 
  properties, and how do Pop III 
  stars differ from later star formation in the presence of metals?}}}

\begin{figure}[th]
\includegraphics[width=.98\columnwidth, angle=0]{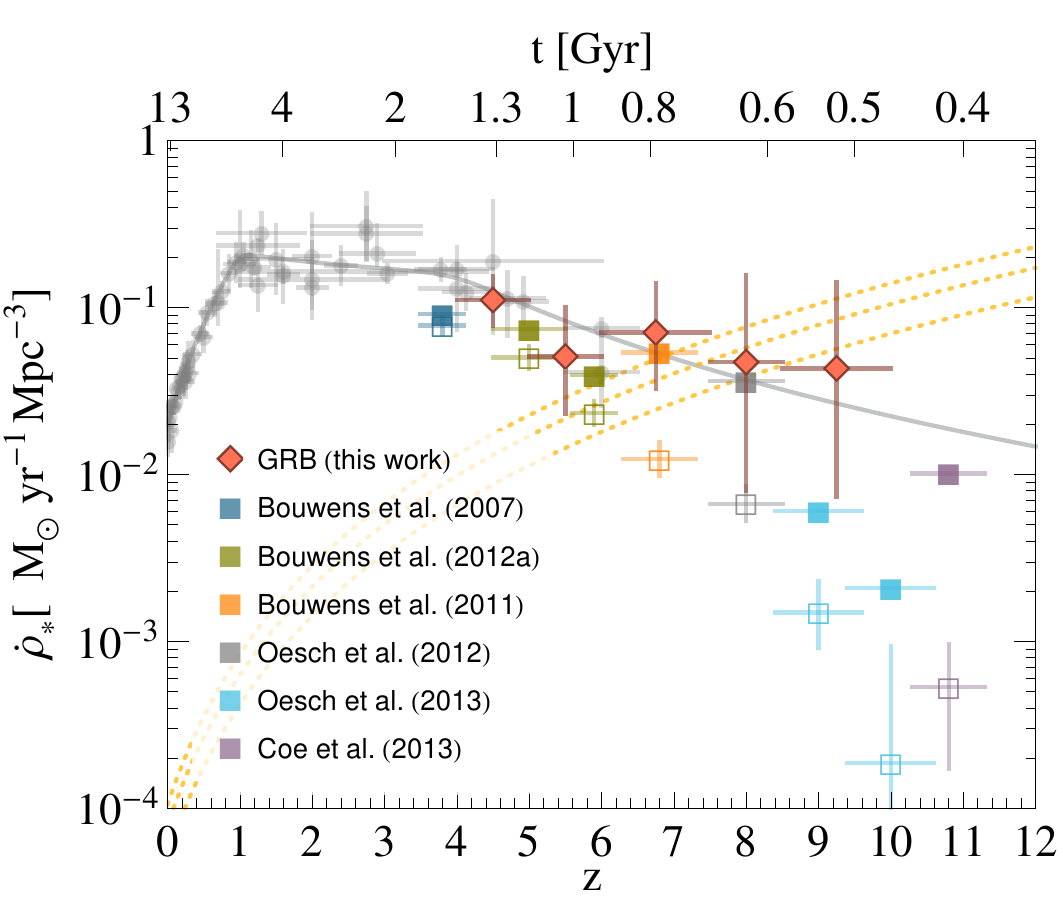}
\vspace{-0.3cm}
\caption[Star formation rate vs. redshift]{
Star formation rate density (SFRD)  
Low-$z$ data ({\it circles}) are from \cite{hob06}.
The {\it diamonds} are obtained using {\em Swift} GRBs.  
The {\it open squares} show the result of integrating the LBG UV luminosity 
functions down to the lowest measured value, $M_{\rm vis}$, while the 
{\it solid  squares} use $M_{\rm cut}\!=\!-10$.  All assume a Salpeter IMF.  
For comparison, the critical $\dot{\rho}_*$ for 
$\mathcal{C}/f_{\rm esc}\!=\!40,\,30,\,20$ (\cite{Madau1999}, 
{\it dotted lines}), top to bottom) are shown.
(From \cite{Kistler2013})
\label{popIII}}
\end{figure}

\subsubsection{Detecting high-$z$ objects}

Direct detections of Pop III or Pop II.5 stars in the early Universe appear 
highly unlikely even with upcoming observatories such as the James Webb Space 
Telescope (JWST) or the proposed 30--40\,m ground-based telescopes 
(such as the E-ELT). Individual stars are much too faint, and only rich 
clusters of very massive stars might be bright enough to lie above the 
detection limits in long exposures (e.g. \cite{jgb09}). 
High-redshift observations seem only able to provide indirect constraints 
on the physical properties (mass, luminosity, frequency, etc.) of the first 
and second generations of stars, for instance, by looking at their influence 
on reionisation or on the cosmic metal enrichment history \cite{CiF05}.

\begin{figure}[ht]
\includegraphics[width=0.49\textwidth, clip]{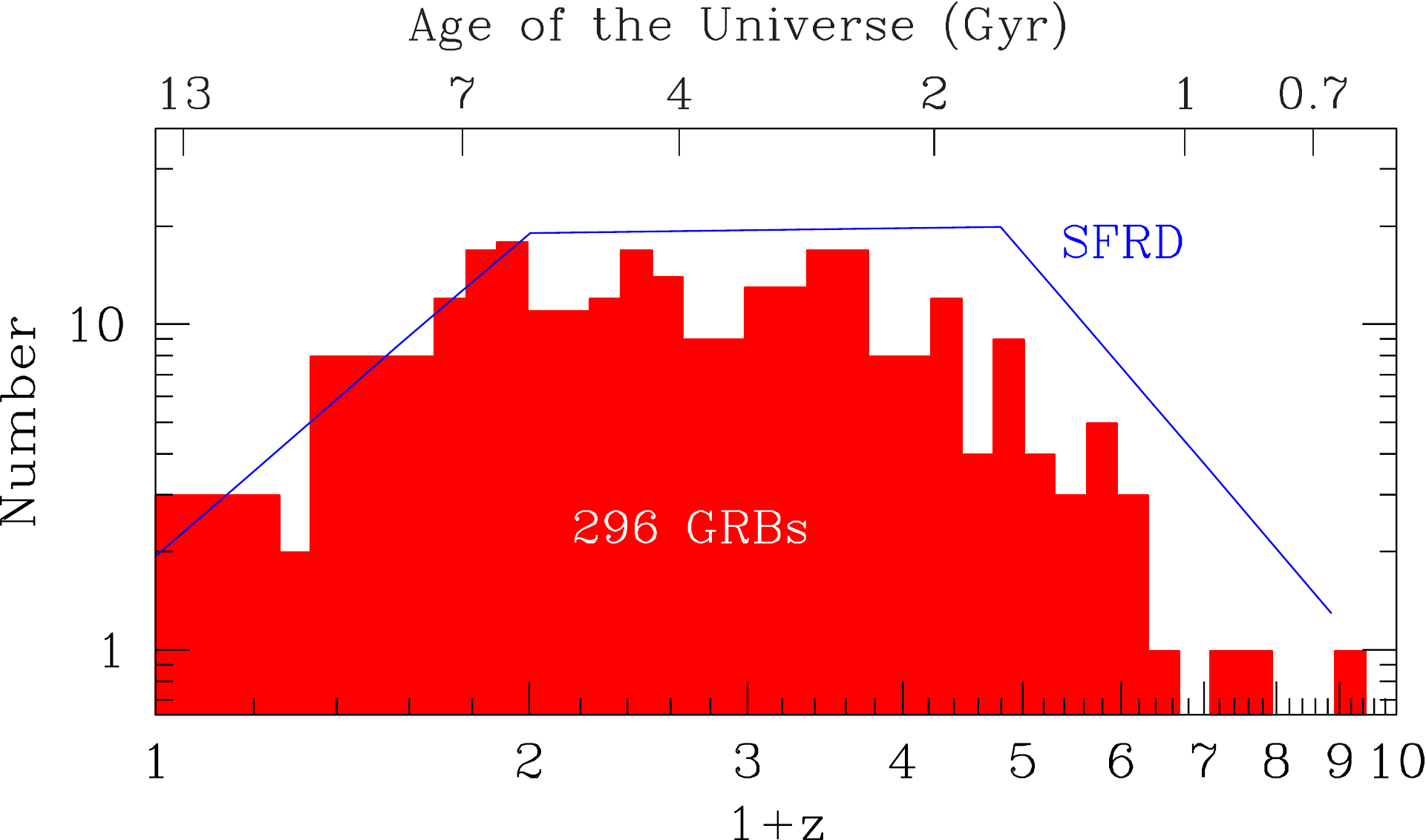}
\vspace{-0.5cm}
\caption{Histogram of the observed number of GRBs (spectroscopic redshift
only) per redshift bin by May 2013 (in
units of $\log (1 + z)$). 
The number of GRBs increases from $z=0$ to $\sim 1$, is steady
up to $z \sim 2.8$, then it decreases down to zero at $z \sim 9$. At low or
high $z$, redshifts are mainly measured front the host galaxy or the DLA
detected in the optical afterglow, respectively. Dust (mainly at $z = 1 -
3.5$) and $gamma$-ray flux detection limits (for $z > 3.5$) affect our 
high-$z$ detections.
This is consistent with the comparison with the SFRD (co-moving volume
change included) derived from field galaxies \cite{hob06}, scaled 
to match the observed $z < 1$ GRB histogram. 
This suggests that a substantial 
fraction of GRBs at high redshift is presently missed.
}
\label{grbsfr}
\end{figure}

The polarisation data of WMAP, the Wilkinson Microwave Anisotropy Probe 
(and likely soon the Planck mission)
indicate a high electron scattering optical depth, hinting that the 
first stars formed 
at high redshift \cite{nab07, bl06, Campisi2011}.
Massive, low-metallicity Pop III stars may produce very
powerful long GRBs \cite{MeR10, KoB10, SuI11}.
Thus, GRBs offer a powerful alternative route (Fig.~\ref{popIII})
to identifying high-$z$ objects, as demonstrated by GRBs 080913 at $z=6.7$ 
\cite{gkf09},
090423 at $z=8.2 $ 
\cite{2009Natur.461.1254T, 2009Natur.461.1258S}
and 090429B at $z=9.4$ \cite{clf11}.
Indeed,  studying GRBs is the only realistic pathway towards the direct 
detection of Pop III and high-mass Pop II.5 stars and thus towards 
constraining their mass spectrum as well as their multiplicity. 
From the predicted mass range of Pop III stars and their high binary 
frequency it was concluded that a $<$0.6\%-2\% fraction of Pop III stars 
ended their lives in GRBs. While at the {\em Swift} sensitivity 
level only $\sim 10\%$ of GRBs detected at $z >6$ could be 
powered by Pop III stars, this fraction increases to 
$40\%$ at $z >10$  \cite{Campisi2011}. 
In addition, both main production channels of GRBs, 
core collapse supernovae of massive stars (long GRBs) as well as binary mergers 
involving Roche-lobe overflow and common-envelope evolution (short GRBs) 
\cite{fwh99}, are likely to be present. This makes high-$z$ 
GRB observations {\em the} ideal probe of studying early star formation
(Fig.~\ref{grbsfr}).

The rate of GRBs is expected to track the global cosmic star-formation 
rate \cite{fps08, KYB09, egk12} (Figs. \ref{popIII}, \ref{grbsfr}),
though possibly with different efficiencies at high-$z$ and low-$z$ 
\cite{Daigne2006, wap10}.
Deduced from a principal component analysis on {\em Swift} GRB data, the
level of star formation activity at $z$=9.4 could have been already as
high as the present-day one \cite{isf11}, a factor 3--5 times higher than 
deduced from high-$z$ galaxy searches through drop-out techniques.
If true, this might
alleviate the longstanding problem of a photon-starving reionisation; it might
also indicate that galaxies accounting for most of the star formation activity
at high redshift go undetected by even the deepest searches. Clearly,
observing more GRBs would be crucial to shrink the currently large error bars
at the highest redshifts, 
thus answering the question:
\colorbox{lightblue}{\parbox{0.98\columnwidth}{
\textcolor{blue}{\bf  What is the relation between GRB rate and star 
  formation rate, and what is its evolution with time?}}}

\medskip

Already with current technology we can characterise GRBs up to redshifts 
of $z \sim 10$ \cite{clf11, 2009Natur.461.1254T, 2009Natur.461.1258S}, 
but reaching larger redshifts requires a new approach and a dedicated mission. 
The present {\em Swift} samples of GRBs, both large biased samples as well as 
smaller but nearly complete samples, indicate a fraction of 5.5$\pm$2.8\% 
GRBs at z$>$5 \cite{gkk11, scv12}. Using standard cosmology and star 
formation history description (Fig.~\ref{popIII}), this translates into a 
fraction of 1\% of all GRBs located at $z>10$, or 0.1\% of all GRBs 
at $z>20$ \cite{egk12}. With 1000 GRBs per year, and a nominal lifetime 
of 5 yr (goal 10 yr) we would expect 50 (goal 100) GRBs at $z>10$, and 
5 GRBs (goal 10) at $z>20$. Thus, the measured GRB redshift count will 
be large enough to observationally constrain the cosmic star formation rate 
at very high redshifts and it will allow us to determine the earliest 
cosmic time when star formation became possible - thus answering the question:
\colorbox{lightblue}{\parbox{0.98\columnwidth}{
\textcolor{blue}{\bf  What is the true redshift distribution and 
corresponding   luminosity function of long-duration GRBs?}}}

\subsubsection{Chemical evolution in the Early Universe}

Beside their direct detection, clues about the first stars can be
obtained by studying the gas polluted by first supernova explosions 
\cite{wbg12}.
Recent models for the formation of Pop III stars suggest
   that their typical masses are similar to those of present-day
   O stars, implying that they will die as standard core-collapse
   supernovae (CCSNe). However, it is also possible that some Pop III
   stars may have much larger masses, of the order of a few     
   hundred solar masses. These stars would die as pair-instability
   supernovae (PISNe), leaving no remnants and producing large  
   quantities of metals and dust  \cite{sfs04}.
The metal abundance ratios produced by CCSNe and 
 PISNe are quite distinct, and hence by measuring their relative
 contributions to the metal enrichment of high-redshift gas,
 we can constrain the form of the Pop III IMF.

GRBs offer a particularly rewarding opportunity to study the physical
conditions of the surrounding medium, in various ways. 
i) The UV radiation of the GRB
and its early afterglow ionise the neutral gas  and destroy most molecules 
and dust grains up to tens of parsecs away.
Interestingly, rotational levels of molecules and metastable states of
existing species (O~{\sc i}, Si~{\sc ii}, Fe~{\sc ii}) are populated by UV 
pumping followed by radiative cascades. 
As the GRB afterglow fades rapidly, recombination prevails and 
the populations of these levels changes on timescales of minutes to hours,
imprinting variable absorption lines in the otherwise flat (synchrotron)
afterglow spectrum. This allows us to measure with unprecedented
accuracy the density, composition and ionisation state of the surrounding ISM
\cite{vel04}.
ii) Other tracers of ionization are molecules forming by the impact of
photons (or cosmic rays) on neutral hydrogen, via the formation of 
H$_{2}^{+}$, which rapidly leads to the production of H$_{3}^{+}$ and
heavier molecules \cite{Black98}. GRBs, provide a good environment to 
induce molecule building processes via ionisation.
iii) The detection of metals through optical absorption lines in the
   highest redshift GRBs (e.g., z=6.3, Fig.~\ref{grbospec}) 
  will allow us  
   to determine whether CCSNe or PISNe are primarily responsible 
   for enriching the gas in these high redshift systems. This has
   important implications for models of the initial stages of 
   reionization \cite{mao98,ChF06, ABS06, mcf11} and the 
   metal enrichment of the IGM \cite{SSR04, tfs07}, 
thus answering the question:
\colorbox{lightblue}{\parbox{0.98\columnwidth}{
\textcolor{blue}{\bf  When and how fast was the Universe enriched with metals?}}}

\subsubsection{The first galaxies}

Identifying objects  beyond $z\sim 7$ has proven extremely difficult.
None of the previously claimed UDF galaxy candidates at $8.5<z<10$
could be confirmed by the deeper multi-$\lambda$ UDF12 campaign 
\cite{Ellis2013} (although new candidates were identified).
Even if found, such galaxies only represent the tip of the iceberg,
in star-formation terms: increasing evidence suggests the bulk of 
early star formation happened in small, low-mass,  and very faint galaxies,
inaccessible to optical/NIR surveys.
This is illustrated by the finding that,
at $z > 5$, six GRB host fields have been observed with deep HST/VLT imaging
\cite{Tanvir2012, Basa2012}
with null detections in all cases. If no dust correction is applied (dust
is not expected to be abundant in the Universe at an age of less than 1 Gyr,
especially in small, low metallicity galaxies), the
UV luminosity limit can be translated into SFR$< 2.5$\,M$_{\odot}$\,yr$^{-1}$ 
\cite{2009ApJ...691..182S}.
Particularly remarkable is the deep $m_{AB} > 30.3$ mag NIR 
limit with {\em HST} of the host galaxy of GRB\,090423 at $z = 8.23$ 
\cite{Tanvir2012},
which gives an incredibly low SFR $< 0.06$\,M$_{\odot}$\,yr$^{-1}$.

\begin{figure}[th]
\vspace{-0.9cm}\hspace{-0.8cm}
\includegraphics[width=1.3\columnwidth, angle=0]{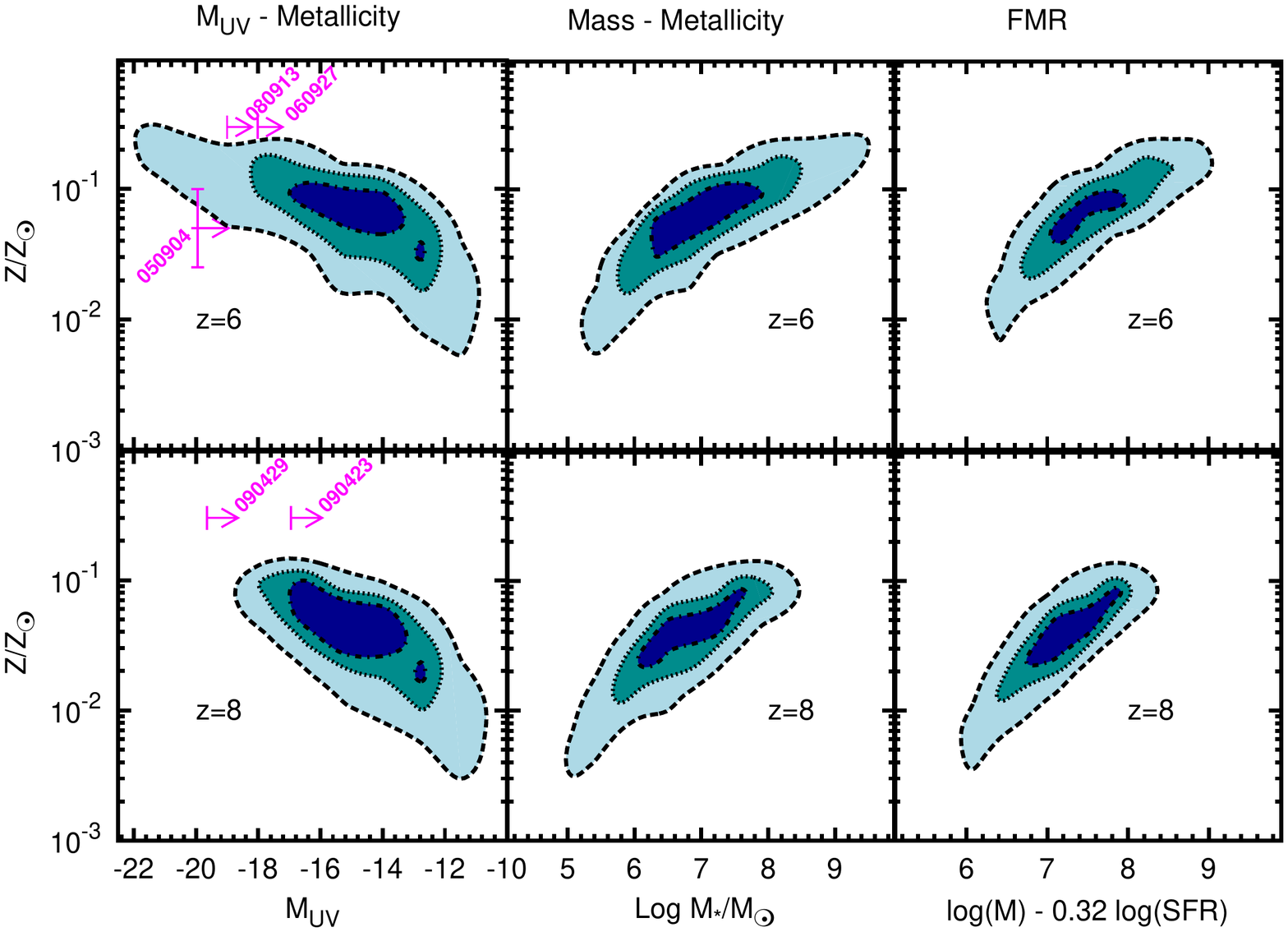}
\vspace{-1.1cm}
\caption{Luminosity-metallicity (left column), mass-metallicity
(central column) and Fundamental Metallicity Relation (right column) for the
LGRB host galaxy simulations at $z=6$  and $z=8$ \cite{smc13}. Contour plots
report the 30\%, 60\%, and 90\% probability of hosting a GRB. Arrows refer to 
\cite{Tanvir2012} and, in the absence of a measured metallicity,   
have been positioned arbitrarily at $Z=0.3 Z_\odot$, while the metallicity of
GRB 050904 has been obtained by \cite{kka06}.
}
\label{grbhosts}
\end{figure}

This finding is in agreement with recent semi-analytic
numerical simulations (Fig.~\ref{grbhosts}) that predict that about 70\%
of GRB hosts at $z > 6$ will be small, with stellar mass in the range 
$M_\star = 10^6$ −- $10^8$\,M$_{\odot}$, while star formation and metallicity 
are in the intervals SFR$= 0.03$−-0.3\,M$_{\odot}$\,yr$^{-1}$ 
and log$Z/Z_{\odot} = 0.01-0.1$, respectively \cite{smc13}. 
For comparison, the deepest rest-frame luminosities
achieved by the HUDF can only reveal down to 
SFR$\sim0.2$\,M$_{\odot}$\,yr$^{-1}$ at $z\sim8$ \cite{bouwens2011}.

Thus, GRBs provide a unique, and above
$z \geq 13$ perhaps the only, way of pin-pointing the vast bulk
of star-forming galaxies as well as their individual
building blocks. 
Furthermore, the
faintness of even the brightest galaxies at $z>8$
makes spectroscopic confirmation very demanding.
GRBs provide the opportunity of probing individual stars
at these times, and their afterglows may provide not only redshifts, but
detailed information about abundances, gas columns etc. via absorption
line spectroscopy.
Indeed, {\em JWST} would be able to obtain $R \sim  3000$ spectroscopy 
at $S/N \sim10$ even 7 days after the GRB explosion, 
while the 30--40\,m ground-based telescopes will be able to provide
unique information on the chemical enrichment and reionisation
history if they can be fed with accurate locations of high-$z$ GRB
afterglows (see Fig.~\ref{EELTspec} for a simulated E-ELT/HARMONI spectrum).

GRB lines-of-sight typically contain more gas than most QSO Damped 
Ly-$\alpha$
systems (DLAs) as they generally probe dense SF regions
within their host galaxies, 
and in that
sense are more representative of high-$z$ star forming environments. 
It required observation of more than 12,000 DLA 
absorbers towards $\sim10^5$ quasars to identify 5 systems with 
log$N_{\rm HI}[{\rm cm}^{-2}]\geq 22$ (0.04\%, \cite{npc12}). 
In contrast, of the 31 DLAs with log$N_{\rm HI}[{\rm cm}^{-2}]\geq 21.4$ 
detected in the GRB afterglow population, 35\% have 
log$N_{\rm HI}[{\rm cm}^{-2}]\geq 22$ 
(e.g. \cite{2006AA...460L..13J, 2009ApJS..185..526F}).

GRBs are already allowing us to see into the heart of star-forming galaxies 
from $z \approx 0$ to $z > 8$
\cite{2009Natur.461.1254T, 2009Natur.461.1258S, 2012ApJ...752...62J}.
With afterglow spectroscopy (throughout the
electromagnetic spectrum from X-rays to the sub-mm) we can characterise 
the properties of star-forming galaxies over cosmic history
in terms of mass function, metallicity, molecular and dust content, ISM
temperature, etc.  Deep follow-up searches for their hosts can then
place strong constraints on the galaxy luminosity function, either through
weak detections (unlike LBG searches this does not require multi-band photometry
for SED fitting), or non-detections which indicate the amount of star formation
in undetectable galaxies.

\subsubsection{Initial stages of re-ionisation}

The reionization of the IGM is the subject of intensive investigation 
currently, and this is likely
to continue for  the foreseeable future.  The fundamental unanswered question 
in the field
is whether radiation from early stars was sufficient to have brought
about this phase change?  If not, then we will be compelled to find
alternative sources of ionising radiation which, given that emission by 
quasars seems to fall well short  of providing the necessary ionising flux 
at $z>3$, may entail new physics
such as decaying particle fields.  Conversely, if early stars are the 
explanation, then
reionization will teach us about their nature and the time-line of their
creation.
At the present time 
it is hard to reconcile the measured star formation
with the required ionising background without invoking, e.g., a high Lyman 
continuum escape fraction,
and/or a dominant contribution from a large population of dwarf galaxies
(as is, in fact, indicated by studies of high-$z$ GRB hosts
\cite{Tanvir2012}), but different from \cite{Kistler2013}.

Various observational windows on the process itself have begun to
produce important results.
Estimates of the optical depth to electron scattering of the cosmic
microwave background (CMB)
by WMAP and Planck indicate a reionization redshift of $z\sim10.4\pm1.2$ 
for an instantaneous reionisation.
From an analysis of 17 ~$z>5$ quasar spectra it was concluded \cite{gff08},
that the HI fraction $x_{\rm HI}$ evolves smoothly from 10$^{-4.4}$
at $z$ = 5.3 to 10$^{-4.2}$ at $z$ = 5.6, with a robust upper limit 
$x_{\rm HI} < 0.36$ at $z$ = 6.3.
However, most limits are model dependent; in fact it was shown that
reionization extending to $z<6$ is not ruled out by current data
\cite{ChF06, mes10, mcg10, ciardi12}.

In the near future, redshifted 21 cm mapping with LOFAR,
MWA and PAPER
are  likely to better establish the timescale of reionisation,
and ultimately much more precisely with SKA.
However, the fine-scale topology of the process, and the key question of 
the nature of the
sources responsible for the ionising radiation will remain uncertain.
GRBs can provide a unique census of early
massive star formation, and a route to understanding the populations of 
galaxies in which they formed.
Crucially, in addition, high-S/N infrared spectroscopy of GRB afterglows 
can provide simultaneous estimates of the neutral hydrogen column
density both in the host \cite{2009ApJS..185..526F} and
in the IGM surrounding it \cite{BaL04, McQ08}
via the shape of the red damping wing of the \lya\ line.
  While most of the flux on the blue side
of \lya\ is simply absorbed for a wide range of neutral fractions,
the shape of the red wing depends on the neutral 
hydrogen fraction of the IGM in which the source is embedded, the host 
neutral column and the extent of any Str\"omgren region around the host 
\cite{Mes04}.
Although this is complicated by the requirement to disentangle
the HI absorption in the host from that in the IGM, in principle,
it can be done as exemplified by GRB 050904 at $z=6.3$ despite a
low-S/N spectrum and high host $N_{\rm HI}$ (Fig.~\ref{grbospec}) \cite{Tot06}.
A large sample of high-$z$ GRBs will likely provide a fraction of
absorbers with low column density, allowing us to cleanly isolate the IGM 
damping wing. The scatter in the 
IGM absorption from an inhomogeneous reionisation is itself a robust
reionisation signature, which can be statistically isolated in a reasonably
large GRB sample \cite{MeF08}.
The exciting prospects for such studies in the era of 30--40\,m ground-based
telescopes is illustrated by the simulation in Fig.~\ref{EELTspec}.

Thus, a sample of a few dozen GRBs at $z>8$ would constrain 
not only the progress of reionisation,
but its variance along different lines of sight (which may be correlated with
identified galaxy populations at the same redshift), and also the typical escape
fractions of radiation from early massive stars.  The latter is a crucial, but
extremely hard to quantify, piece of the puzzle, since only if the ionising 
radiation can escape unimpeded from a significant fraction of massive stars 
(say, $>20$\%),
will they be successful in driving reionization.  Measuring directly the 
neutral columns
to many GRBs will establish how many lines of sight provide such an unabsorbed 
view.
With a fiducial GRB-finder with 1000 GRBs/yr and immediate redshift 
estimates, ground-based spectroscopy can be secured for many dozen
GRBs in the $6<z<13$ range.

An unique and independent way of probing the high-$z$ UV radiation 
field with GRBs is through its effect on high-energy photons.
The expected UV field at these redshifts can
cause appreciable attenuation above a few GeV, that can be observable with 
e.g. CTA \cite{Ino10}.

In conclusion, a powerful GRB detection and localisation mission,
in tandem with future facilities expected to be available on a 15--20 year
time frame, will answer the question:

\colorbox{lightblue}{\parbox{0.98\columnwidth}{
\textcolor{blue}{\bf How did reionisation proceed as a function of 
environment, and was radiation from massive stars its primary driver?}}}

\subsubsection{Warm-hot IGM studies}

The redshift distribution of X-ray absorbing column densities, $N_{\rm HI}$,
as detected in GRB afterglows by {\it Swift}/XRT shows a significant
excess of high $N_{\rm HI}$ values at redshifts $z\ge 2$
 with respect to the low redshift GRBs \cite{Camp10, Camp12, WaJ12}.
This excess absorption has been tentatively interpreted as due to 
the presence of absorbing matter along the line of sight not related to 
the GRB host galaxy. This can be either diffuse (i.e. located in diffuse
structures like the filaments of the Warm-Hot Intergalactic Medium -
WHIM \cite{Behar11, Star13})
or concentrated into 
intervening systems (i.e. galaxies or clouds along the line of sight
\cite{Camp10, Camp12}. The study of X-ray absorptions for a larger
sample of GRBs at redshifts $z\ge 2$ could provide new insight on
the nature of the intergalactic medium and in particular allowing to
constrain its metal content.

Quasars are the alternative target for this kind of studies, in fact
WHIM signatures have been detected when observing the bright blazar Mkn
421 (e.g. 
\cite{Nic05}). GRBs provide a much larger flux, if
observed promptly, allowing us to extend these studies to larger
distances. It is not easy to disentangle filaments from intervening
systems, whereas a sufficient spectral resolution will allow us to 
detect distinct absorption features (originating at a given redshift)
versus a truly diffuse medium (across a redshift range). As a by-product
a direct measurement of the GRB
redshift can be obtained from the X-ray data alone \cite{Camp11}.

\subsection{The GRB origin}

\subsubsection{GRBs and neutrinos }

Neutrinos  are  electrically  neutral, weakly  interacting  elementary
particles  which are  produced as  the result  of  radioactive decay,
nuclear  reactions or  proton-proton collisions. Examples are  the fusion
reaction in  the Sun,  electron capture during  the collapse  of stars
into a supernova, and  particle acceleration (jets) in e.g., active
galactic  nuclei,  microquasars, supernova remnants or GRBs \cite{becker08}.  
Due  to  the  very  small
interaction cross section, neutrinos  are difficult to detect, but the
detection  sensitivity increases  dramatically  with neutrino  energy.
This disadvantage is however an advantage for the search for neutrinos at the
same time: they are neither absorbed nor deflected on their way to Earth,
so that the production region can be studied.
This makes them unique in the search for the origin of ultra-high energy 
cosmic rays.

The vast majority of the neutrinos from a GRB is emitted at moderate
energies ($\sim$ 20 MeV) from the central engine's accretion disk.  
Their moderate energies together 
with the steep energy dependence of the interaction cross sections make
them hard to detect.
Chances are much better for those neutrinos produced by the
ultra-relativistic outflow that is responsible
for the GRB prompt emission. 
The   GRB  fireball   phenomenology
predicts spatially and  temporally correlated neutrino emission 
to occur from proton-proton or proton-photon interaction.  
For a  neutrino  
flux  distributed as  a  power law  $\propto
E^{-2}$,  this implies that energies in the range TeV to PeV  are  
most promising
for neutrino  detection from distant  sources \cite{WaB97}.

A number of possible neutrino production sites from long GRBs have been
identified: within the exploding star, within the relativistic outflow, and 
within the reverse shock that is formed as the 
afterglow is developing. Neutrinos can be
formed in proton-proton and proton-photon
interactions in the jet cavity that is formed as the jet penetrates 
the collapsing star. This is expected to produce
a flash  of neutrinos  with energies  of $3-10$\,TeV.
Alternatively, neutrinos can be produced  in the same region as the
$\gamma$-ray  photons,  within the  jet.   Here,  the
so-called  prompt neutrino  emission with  energies  of $\sim100$\,TeV
should accompany the $\gamma$-rays.
The detailed timing of neutrino and $\gamma$-ray emission can constrain
the physics of the GRB emission.

Despite sophisticated  searches, neutrinos from GRBs have not been detected  
so far.  While our best hope for
neutrino detection is with the continued operation of IceCube until (at least)
2020, the follow-up project KM3NeT is in its extended design phase,
with the implementation of the first phase of the infrastructure being 
immanent. The neutrino detection from GRBs would clarify the hadronic
content in GRB jets. Moreover, systematic
measurements of the neutrino energies, in particular if they peak at
certain key energies, could help discriminate between models, even more so 
when combined with properties of the measured $\gamma$-ray spectrum. 
A neutrino detection from GRBs would also directly prove 
GRBs as sources of ultra-high energy cosmic rays.
The ratio of neutrinos to $\gamma$-rays, typically produced in 
similar numbers, would provide indications, otherwise difficult to obtain, 
on the attenuation of $\gamma$-rays in the early stages of the fireball.

Detection of neutrinos from cosmologically remote GRBs 
(i) provides limits on the lifetime of the 
  dominant mass eigenstate by a factor $>$200 better than for SN 1987A;
(ii) is a testbed of neutrino properties 
with an unprecedented accuracy; 
(iii) tests if neutrinos follow the weak
equivalence principle;
(iv) facilitates the
exploration of quantum-gravity-induced Lorentz invariance violation;
(v) provides tremendous advantage over other methods of studying cosmology,
as neutrino flavor ratios should be independent of any evolutionary effects.

In addition to those microphysics-related goals, the detection of
high-energy neutrinos from GRBs aims at answering the astrophysical 
questions of the (i) identification of the sources of ultra-high energy   
cosmic rays; (ii) determination of the ratio of accelerated electrons  
to protons in GRBs, (iii)  proper treatment of the GRB jet physics,  
including hadronic cosmic rays. 

In order to achieve those central goals, neutrino telescopes rely heavily on 
satellites that trigger GRBs: neutrino analyses can improve their
sensitivity by reducing the main background of atmospheric neutrinos to
almost zero through the selection of events in space and time, according
to the occurrence of GRBs. This makes the GRB analysis one of the most
sensitive ones for cosmic neutrinos. Only with existing
satellite triggers,  we can answer the question:

\colorbox{lightblue}{\parbox{0.98\columnwidth}{
\textcolor{blue}{\bf How are $\gamma$-ray and neutrino
  flux in GRBs related, and how do neutrinos from long GRBs constrain the 
  progenitor and core-collapse models?}}}

\subsubsection{GRBs and gravitational waves}

Short GRBs (sGRBs) and GWs are linked by the common topic ``compact binary
mergers'' and they nicely illustrate how complementary and mutually beneficial
the information obtained in both channels is \cite{bloom09, phinney09}. 
Moreover, the additional EM signals expected from a compact binary merger 
provide a close link to cosmic nucleosynthesis.

About one quarter of the {\it CGRO}/BATSE and {\it Fermi}/GBM
bursts are classified as short-duration ($<2$\,s), hard GRBs. 
As short GRBs are 
intrinsically less luminous in EM radiation than their long-duration cousins, 
the observed sample is dominated by relatively nearby sources. The presently 
known redshift distribution suggests that a detection rate 
of 1000 GRBs/yr corresponds to 10--15 short GRBs/yr at $z<0.1$ ($\sim$450\,Mpc) 
 \cite{Siellez2013, Coward2012}, depending on the energy 
range of the trigger instrument.

The question of their central engine is a long-standing puzzle. Compact
binary mergers (either NS-NS or NS-BH) are the prime suspects, but this 
connection is far from proven. The coincident detection of a sGRB and
a GW  signal could finally settle this issue. 
The network of the gravitational detectors Advanced {\em LIGO/VIRGO}, 
soon complemented 
by LIGO-India and KAGRA is expected to deliver the first direct GW detections
within a few years from now. It will be capable of identifying an
optimally oriented NS-NS (NS-BH) merger out to $\sim$450 ($\sim$900) Mpc,
with a combined predicted rate of the order of 50 yr$^{-1}$ \cite{aba10}.
The GW signal of a compact binary merger potentially delivers a wealth of
information on the physical parameters of the binary system. For example, it
provides the 
neutron star masses and radii, it carries the 
imprint of the equation of state at supra-nuclear densities, and it
constrains the collapse stages, e.g., through a hypermassive NS or 
magnetar to a BH, information that is hardly accessible otherwise. 
Comparison of the rates of GW detections with and without sGRB 
counterparts may 
constrain the 
geometry of the relativistic outflow (``jet''), the source energetics
and the physical emission processes. But while providing a clear view
on the physics of the actually merging system the poor localizations 
by GWs of $\approx$10--1000 square degrees \cite{Aasi2013, kli11} leave 
us nearly blind with respect to the astrophysical environment in 
which the merger takes place. 

A complementary EM detection can provide a wealth of additional 
information. Firstly, it locates the source for optical follow-up providing 
an accurate localization relative to the host galaxy, thus allowing 
us to study the environment of such evolved sources. This, in turn, 
constrains binary stellar evolution by providing information on kick velocities,
initial separations etc. 
Secondly, a redshift and thus luminosity determination combined with the 
absolute source luminosity distance provided by the GW signal can deliver 
precise measurements of the Hubble parameter (10 GW+EM events in Advanced 
LIGO/Virgo may constrain the Hubble parameter to 2-3\% \cite{dalal06}, 
and ET will constrain it to $<$1\% \cite{Sathy10}), and hence help to break the 
degeneracies in determining other cosmological parameters via CMB, 
SN\,Ia or BAO surveys.
Thirdly, the detection of a radioactively powered transient \cite{LiP98}
may provide an interesting 
link to cosmic nucleosynthesis: this could show the ``r-process in action'' 
and finally settle the question of where the heaviest elements 
around the platinum peak (nucleon numbers $A \sim 195$) come from. 
Neutrino-driven winds from a merger remnant \cite{Dessart09}
may lead to yet another radioactive transient, but with likely different 
properties. Once the matter that is dynamically ejected interacts with 
the ambient medium it may produce radio flares which independently 
would set a limit on the merger rate \cite{pnr13}.

The localisation of GW events has another more subtle benefit: it 
improves the accuracy with which parameters can be estimated from the
GW observation \cite{SaSchu2009}. The covariance of
angular errors with uncertainties in other parameters (distance, polarisation,
stellar masses etc) is usually significant. Thus, a more accurate position
through EM follow-up also improves the determination of all the
parameters measured gravitationally.
For short GRBs, several of the GW events will be near threshold, and because 
the GW amplitude is peaked along the jet axis, the detection range 
increases by a factor of $\sim$2 with coincident detection of a short GRB 
X-ray afterglow \cite{cut02}.

Only with a sensitive GRB detector in orbit, operating in conjunction
with the gravitational wave detectors, can we answer the question:
\colorbox{lightblue}{\parbox{0.98\columnwidth}{
\textcolor{blue}{\bf Can short GRBs be unambiguously linked to 
  gravitational wave signals, and what do they tell us about the neutron 
  star merger scenario?}}}

\subsubsection{ Gamma-ray polarisation}

Until 5 years ago,
the  prompt 20--1000\,keV  emission was
interpreted as  a smoothly  broken power law produced  by synchrotron
emission.  Recent  discoveries of additional  spectral components at
high  and low  energies with {\it Fermi}, as well as 
$\gamma$-ray polarisation measurements with {\em INTEGRAL} and {\em IKAROS}
have dramatically challenged our view of the GRB emission process.
 Is the broken  power law a Comptonised thermal component from
  the  photosphere?   Is  the  high-energy part  produced  by  inverse
  Compton  radiation  and  the  low-energy  component  of  synchrotron
  origin?   

Time-resolved $\gamma$-ray polarimetry 
of the GRB  prompt  emission would be a unique discriminant of the underlying
physics. The level of polarisation will depend on the radiation 
mechanism as well as geometrical effects. In particular, it will probe 
the  strength and scale of the magnetic field.
A significant level of  polarisation can be produced by
either  synchrotron emission  or  by inverse  Compton scattering.  The
fractional  polarisation from  synchrotron   emission  in  a
perfectly  aligned magnetic  field can  be as  high as  70--75\,\% 
\cite{Granot03, Toma09}. An
ordered magnetic  field of this type  would not be  produced in shocks
but could be advected from the central engine \cite{Granot03, gk03}.
Strong correlations are predicted
between the polarisation level, the jet Lorentz factor and the power-law index
of the particle distribution \cite{lyu03}.
Another asymmetry  capable of
producing  polarisation,  comparable  to  an ordered  magnetic  field,
involves  a jet with  a small  opening angle  that is  viewed slightly
off-axis \cite{waxman03}.
In the case of photospheric emission, as recently hotly debated based
on {\em Fermi} data, polarisation can arise due to the multiple Compton
scatterings before photons escape \cite{Belo11}.
Measurements of the temporal evolution of both, the degree of polarisation
as well as the polarisation angle have strong diagnostic power to constrain
GRB models.

Recently some measurements of polarisation during the prompt emission of GRBs 
in the hundreds of keV energy range have been reported 
\cite{kalemci07,McGlynn07,mcglynn09,gotz09,yonetoku11,yonetoku12,gotz13}. 
Although all these measures, taken individually, have not a very high 
significance ($\gsim$3 $\sigma$), they indicate that GRBs may indeed be 
emitters of polarised radiation. In particular, the changing polarisation angle
with time \cite{gotz09, yonetoku12} indicate 
a fragmented jet. This kind of polarisation measurements 
can shed new light on the strength and scale of magnetic fields, as well as 
on the radiative mechanisms at work during the GRB prompt emission phase.

In addition, polarisation measures in cosmological sources are also a 
powerful tool to constrain Lorentz Invariance Violation (LIV), arising 
from the phenomenon of vacuum birefringence as shown recently
\cite{gotz13, fan07, laurent11a, toma12}.

The next generation of instruments will be sensitive enough
to not only provide averaged  polarisation   angles  and  degrees  for  each
detected event (long and short bursts), but even more pulse-resolved 
measurements for the brighter events. 
The detailed analysis of the prompt
emission polarisation properties in GRBs would lead to essential clues to the
emission mechanism. In particular, an ordered magnetic field can be determined
or ruled out.

\subsection{Time-domain astrophysics}

It is now widely accepted that the next astronomical discovery frontier 
is the time domain (as emphasised in the Astronet Roadmap and in the 
US Decadal Survey).
Current time-domain experiments 
are extremely successful and the
coming years will see a revolution in time-domain astronomy with many surveys
in the optical 
and in the radio.

\subsubsection{Other high-energy transient types}

Besides GRBs, also other transient source classes can trigger 
instruments surveying for GRBs.
Transient high-energy sources,
watched in real-time, offer insight into the physics of
accretion, the presence (and mass) of BH in galaxies, 
and the behaviour of matter under extreme gravitational
and magnetic fields, to name but a few. While much science
in these diverse subjects arises from detailed follow-up across
the EM spectrum, many of the events are most dramatic
at higher energies, and hence require high-energy triggers to identify,
even in the era of LSST.

Within the Milky Way our proposed mission will be sensitive to
emission from M-dwarf stars, mapping out the frequency of their
activity and the implications for planet habitability (especially
important as many next generation planet searches are targeting  
M-dwarfs due to improved contrast). We will pinpoint soft gamma-repeaters
-- highly magnetised neutron stars that are possible gravitational waves 
sources 
and which provide a test bed for
physics in both strong gravitational and magnetic fields, and for models
of the supernovae that may create them. Outbursts from X-ray binaries
of various types are also likely to be discovered, potentially even from
outside the Milky Way. 

The breakout of the SN shock from the star might provide a short lived,
but luminous X-ray burst, that has likely been observed in at least
one case (SN 2008D). More generally, SN can create powerful  
X-rays via their interaction with the circumstellar medium, offering
a route to studying mass loss in the years before the stars demise.
X-rays might also be generated from engine-like events deep within
the ejecta that become visible at late times as the ejecta becomes
optically thin. 
Of particular importance is the nature of the superluminous SN
\cite{GalYam12}, whose origin may be similar to the
dominant mechanism thought to operate in Pop III stars, and
which have recently been claimed to be (at least occasionally)
powerful X-ray emitters.

Moving further out into the Universe we can study more
massive black holes in galactic nuclei. The recent discovery of
hard high-energy emission from Swift J1644+57 \cite{bloom11}
suggests that tidal disruption flares (TDFs) 
might be powerful high-energy transients,
while it is also thought that all TDF produce softer thermal
X-ray emission \cite{kom12}.  TDFs offer a unique route to probing BHs
in galaxies, including their location and ubiquity within dwarf  
galaxies (where it is far from obvious they reside), hence they
allow us to extend the relation between BH mass and stellar 
velocity-dispersion to much lower masses,
providing strong constraints on galaxy evolution models.  Finally,
these events allow us to study accretion around supermassive BHs
from switch on to switch-off in human timescales, much shorter
than the millions of years in which active galaxies evolve.

\subsubsection{Complementarity with other transient detection systems}

The main reason for the community-wide focus on the transient and dynamic
Universe is that it is most often associated with extreme physical phenomena:
eruptions on a stellar surface, complete explosion of a star, "shredding" of a
star by a supermassive black hole, merger of two extremely compact objects,
etc. These phenomena most often emit non-EM signals, in  
particular cosmic rays, neutrinos and gravitational waves. 
Our proposed mission concept  
would detect and localise energetic phenomena that are most likely to be 
associated with    non-EM signals.  

By definition, the transient sky is unpredictable which is why all EM facilities
have a very large field of view;  the need for very wide area coverage cannot 
be overstated, and it is crucial to have an EM monitor that sees a large
fraction of the sky all the time. We can do this with a dedicated $\gamma$-ray
mission. Focussing on one wavelength range or one information carrier (e.g., EM,
GW, $\nu$) is like having a black-and-white picture: there is useful 
information but we are missing something. A range of instruments covering the
whole EM spectrum in conjunction with other information carriers will give us a
detailed \textit{color} image, allowing us to see the whole physical picture,
thus addressing the question:
\colorbox{lightblue}{\parbox{0.98\columnwidth}{
\textcolor{blue}{\bf What are the electromagnetic counterparts to 
  gravitational waves and neutrino bursts?}}}

\clearpage

\section{Requirements for enabling instruments}

Both, the use of GRBs as a tool as well as the simultaneous detection of an
EM signal with a GW/neutrino signal, requires an in-orbit trigger and 
search facility (``GRB-finder'') that can 
simultaneously localise the event within the large error boxes 
provided by the GW (and neutrino) facilities.

In order to use GRBs as a tool, positions  with  arcsec accuracy  are  required.
The localization  accuracy of the GRB-Finder will not be sufficient,
and thus an X-ray and/or optical/(N)IR telescope is required which is 
rapidly slewed to the position determined by the GRB-Finder.
An X-ray telescope is preferred since the sky is too crowded at optical/(N)IR
wavelengths.

Finally, to tackle the early Universe questions and obtain decent statistics
at $z>10$, a next-generation GRB mission should detect of order 1000 GRBs/yr,
providing 50 (5) GRBs at $z>10 (20)$ over a 5 yr mission lifetime.
This high GRB rate requires a pre-selection
of 'interesting' events, and therefore a (N)IR telescope is foreseen which
will determine redshifts for  the bulk  of the  high-redshift (e.g., $z>7$) 
sources. Table 1 summarises these high-level requirements.

\begin{table*}[thb]
\caption{Scientific requirements for a future GRB mission with assumed 5 yr lifetime.}
\vspace{-0.3cm}\hspace{-0.5cm}
\begin{tabular}{lll}
  \hline
  \noalign{\smallskip}
    ~~~~Requirement & ~~~~~~~~~~~~~~~~Goal & ~~~~~~~~Detector ability \\
  \noalign{\smallskip}
  \hline
  \noalign{\smallskip}
  $\!\!\!\!\!$1. Detect 1000 GRBs/yr & obtain  50 (5) GRBs at $z>10 (20)$ & 
           large FOV, soft response\\
 $\!\!\!\!\!$2. Rapid transmission to ground & allow timely follow-up observations & communication network\\
  $\!\!\!\!\!$3. Rapid localization to few $\asec$ & opt/NIR identification of 1000 GRBs/yr & slewing X-ray or opt/NIR telescope\\
  $\!\!\!\!\!$4. Provide $z$-indication & allow selection of high-$z$ objects & multi-filter or spectroscopic capability$\!\!\!\!\!$\\
  \noalign{\smallskip}
  \hline
   \end{tabular}
   \label{requir}
\end{table*}

\subsection{The GRB-finder}

The localisation accuracy and timeliness are the crucial parameters 
when aiming at follow-up observations at longer wavelengths.  
We discuss in the following only concepts which provide
localisations better than a few degrees within minutes after the GRB.
In general, as the prompt GRB spectral slope is $-1$ in the 1--100 keV
band, lowering the energy threshold allows for the detection of
a larger number of GRBs.

\paragraph{Scintillation detectors:}

The use of 
simple scintillator detectors like {\em CGRO/BATSE} or {\em Fermi/GBM}
has, in the past, only led to afterglow identifications for a handful of GRBs,
due to their large localisation uncertainties.
The systematic error for {\em GBM} bursts is 3\fdg3 for $\sim$90\% of
the cases, with a tail of 12\fdg5 for the rest.
The twelve NaI detectors on {\em Fermi/GBM} work in the 8--1000 keV band,
and with an effective area of about 100 cm$^2$ each in the 20--50 keV band
they detect $\sim$270 GRBs per year \cite{mlb09}.
Increasing the rate to our fiducial 1000 GRBs/yr can be achieved in 
different ways. Firstly, if flown in, e.g., a L2 orbit, the lack of
the Earth occulting half the sky implies doubling the detection rate.
Secondly, increasing the effective area by simply using larger crystals
is straightforward. Scaling the background rate appropriately and
assuming the same S/N ratio for triggering, an effective area of
10$\times$ that of GBM would provide 1000 GRBs/yr in a low-Earth orbit
(LEO).

\paragraph{Coded Mask Instrument:} 

Such systems have been widely used in space for 
detection of GRBs (e.g. {\em Swift/BAT}), and work in the $\sim$2--200 keV band.
Their advantages are:
i) observe over a large solid angle; 
ii) can use hard X-ray/$\gamma$-ray detectors to cover quite large energy 
bandpass; 
iii) can give fairly good localisations (one to few arcminutes);
iv) provide a large number of photons, allowing easier spectral sanalysis.
Disadvantages are: i) they are non-focussing, so sky background prohibits 
the detection of faint sources or monitoring of fading emission from sources 
which trigger the instrument (i.e. this requires an additional 
focussing telescope which can create a data gap -- as in the case of Swift
-- while the satellite slews); 
ii) while coded mask instruments can be used with large-area Si detectors 
to cover the X-ray band, they have a modest bandwidth (2--50 keV).

Simulations using the presently known log$N$-log$S$ relation 
and luminosity function of GRBs \cite{wap10}
reveal the following trade between depth and area of a coded-mask 
similar to {\em Swift/BAT}: aiming at 2 (4) times the depth of {\em BAT}
gives a similar number of high-redshift GRBs as increasing the detector area
by a factor of 2.5 (5).
In order to achieve $\sim$1000 GRBs/yr, a system of seven {\em BAT}-like 
systems 
with only modestly increased (1.4$\times$) effective area would be necessary
(or correspondingly enlarged versions of the advanced coded mask instruments
{\em SVOM/ECLAIRs} or {\em UFFO/UBAT}). 

\paragraph{Lobster Optic Instrument:} 

The use of a wide-field Lobster Eye (LE) Microchannel Plate (MCP) 
or Multi-Foil (MFO; \cite{hpm11, tichy11})
imaging instrument provides several advantages over traditional coded mask 
wide-field telescopes:
i) one gets continuous monitoring in a single 
bandpass (i.e., no gaps due to slews) as the same telescope finds and 
then continues to monitor the transient; 
ii) the use of a focussing optic lowers the sky background against which 
sources are detected, increasing the sensitivity by about two orders of 
magnitude;
iii)  ability for good localisation ($<$1\amin\, 
down to about 10\asec) particularly for higher focal lengths; 
iv) multiple, lightweight modules can be utilised to cover large solid angles. 
The principle disadvantage is the need of (modular) large area 
detectors (as for coded mask telescopes). LE instruments are
restricted to low energies (of order 0.5--10 keV).
At similar mask/optics area and FOV, a Lobster optics would detect
about 3--4$\times$ more GRBs than a coded mask system \cite{bfp12}.
A detection rate of 1000 GRBs/yr could be reached with about 10 modules of
the type proposed for the Lobster mission \cite{Gehr12}.

\paragraph{Compton Telescope:}

A Compton telescope would work at higher energies 
($\sim$200 keV to $\sim$50 MeV), and has the advantages of 
i) uniquely excellent gamma-ray polarisation capability, and
ii) a wide energy band. The disadvantage is a localisation 
accuracy 
substantially poorer than a coded mask or Lobster optics instrument, 
of order 1\degs\ radius only, and a rather large mass.
An existing concept of such a detector promises $\sim$600 GRBs/yr 
\cite{grips2012}, close to our 1000 GRBs/yr goal.

\subsection{The X-ray telescope for precise localisation (and spectroscopy)}

\begin{figure}[ht]
\includegraphics[width=0.99\columnwidth]{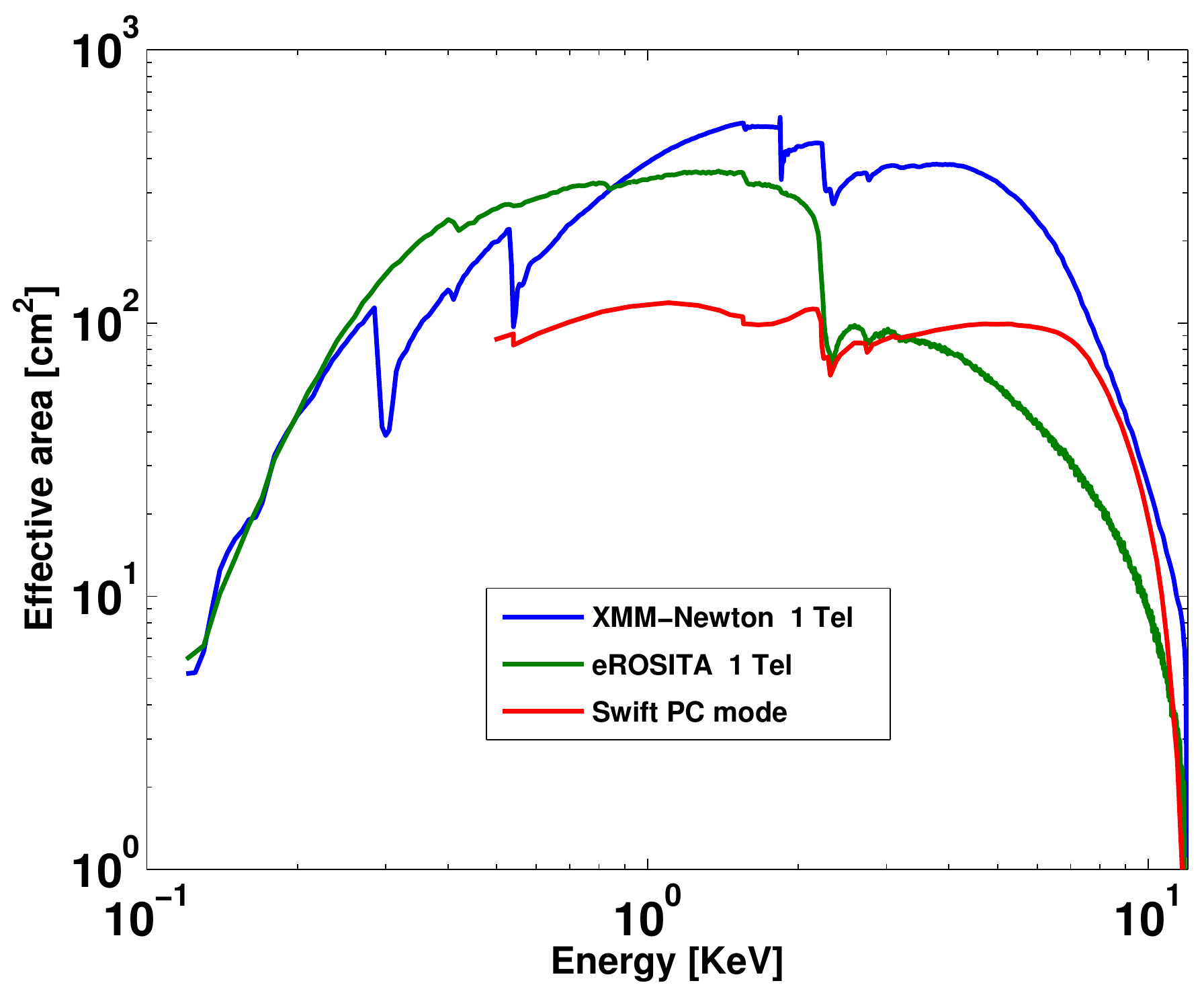}
\vspace{-0.2cm}
\caption[XRT effective area comparison]{Comparison of the effective
area of a modified eROSITA system (one telescope per sky position
instead of all 7 telescopes co-aligned)
with those of XMM and Swift/XRT. (From \cite{grips2012})
\label{xrteffarea}}
\end{figure}

The main driver for the design of the X-ray telescope (XRT) is the 
position uncertainty provided by the GRB-finder, such that the full GRB 
error circle can be covered. In addition, 
the sensitivity should allow all GRB afterglows
to be detected. Scaling from the {\em Swift/XRT} detections of
the faintest GRBs, and considering the goal of reaching substantially
larger redshifts (and thus likely fainter afterglows), the XRT should 
be a factor $\sim$3 more sensitive than {\em Swift/XRT} (Fig.~\ref{xrteffarea}).
Such sensitivity requirement (of order 
10$^{-13}$ erg/cm$^2$/s in 100 sec)
excludes coded mask systems. 

In case of a Compton telescope as GRB-finder, a FOV of 3\degs\ diameter is
needed. Combined with the sensitivity requirement, a single-telescope 
Wolter-I optics is problematic. A practical solution is to
adopt the {\em eROSITA} scheme of 7  Wolter-I telescopes,
and adjust their orientation on the sky such that they fill 
the required FOV.
For the other two GRB-finder options a single {\em eROSITA} telescope
would be sufficient, or alternatively the {\em XMM} flight spare (though larger
and more massive).
Simpler versions like an enlarged version of the {\em SVOM/MXT}
or a long focal-length Lobster are possible as well, with the trade-off
of less versatile auxiliary science options as compared to {\em Swift/XRT}.

\subsection{The (N)IR telescope}

The main driver for the design of the Infrared Telescope (IRT) is the 
goal to (i) detect and accurately localise the counterpart, and
(ii) measure the redshift to an accuracy of at least $\Delta z / z \sim 0.2$,
so that high-$z$ GRBs can be quickly identified for detailed follow-up study.
Above redshift $z \sim 17$, \lya\ is moving out of the $K$-band.
This and the requirement to be sensitive up to redshifts of $\sim$30
imply to observe in the $L$ (3.5 $\mu$m) and $M$(4.5 $\mu$m) bands.

\begin{figure}[ht]
\vspace{-0.4cm}
\hspace{-0.7cm}\includegraphics[angle=270,width=1.2\columnwidth]{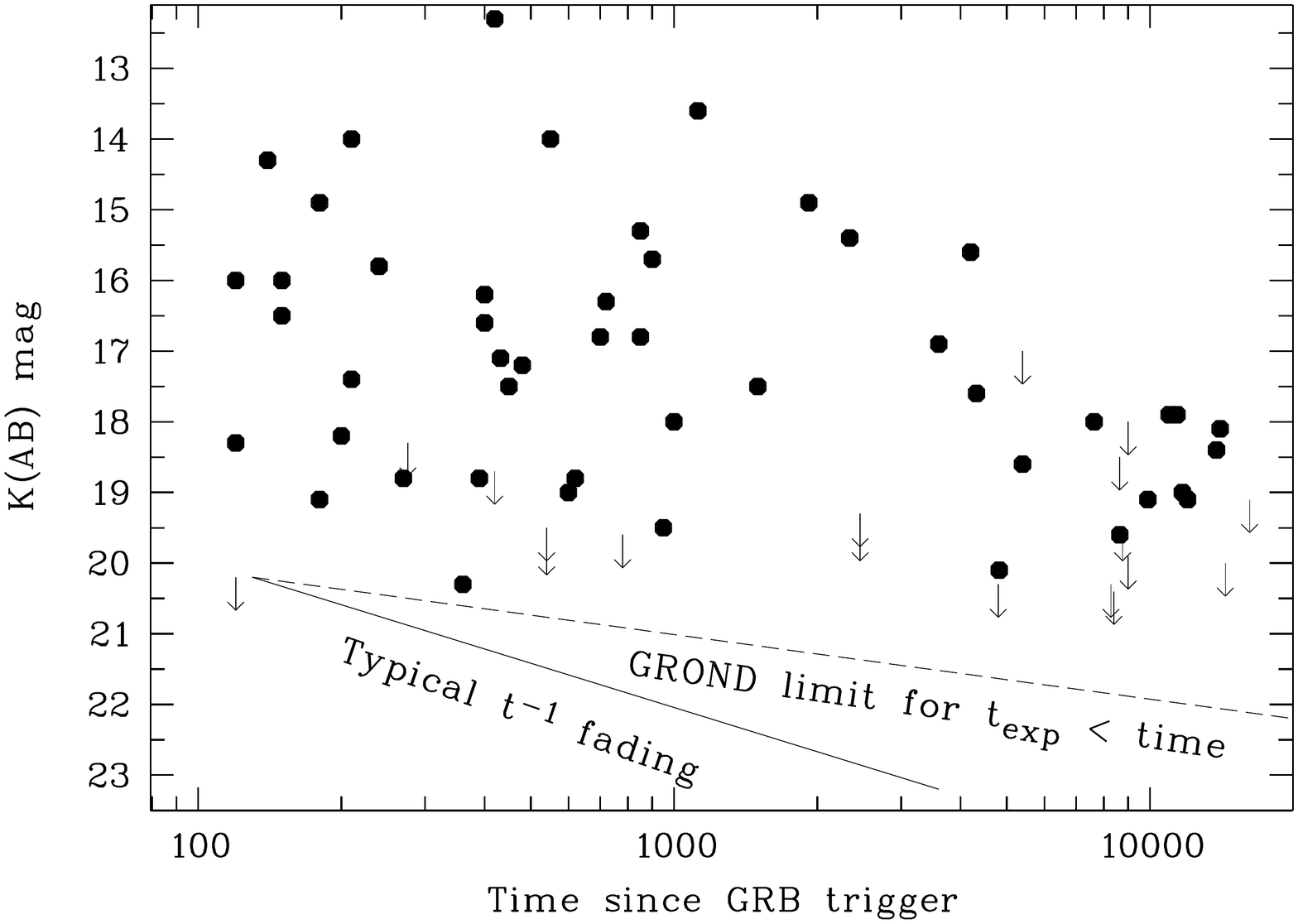}
\vspace{-0.6cm}
\caption{$K$-band photometry of a complete sample of GRB afterglows
based on GROND data.
At 1000 s after the GRB, 95\% of the afterglows are brighter
than $K$(AB)=22 mag. With $K-M \sim 0.8$ mag for typical afterglow
spectral slopes, we aim at $M$(AB)=21.2 mag at 1000 sec, or
$M$(AB)=22.2 mag at 1 hr after the GRB.
}
\label{IRTlimit}
\end{figure}

Based on a complete sample of GRB afterglow measurements obtained with
the 7-channel optical/NIR imager GROND since 2007 (update of \cite{gkk11}), 
in particular the 
brightness distribution in each of the JHK channels, a minimum afterglow 
brightness 
of $M$(AB) $\approx 22$ mag at $\sim$1 h after the GRB is deduced 
(Fig.~\ref{IRTlimit}).
Such sensitivity will be reached with the future 30\,m class telescopes,
but since it is illusory to follow-up 3 GRBs/night with those instruments,
we consider this an onboard requirement in the following.

Using standard parameters for the transmission of the optical components,
read-out noise of the detector as well as zodiacal background light,
a 1\,m class telescope would achieve at least a 5$\sigma$ $M$-band detection 
of each GRB afterglow with a 500 sec exposure when observed within
2 h of the GRB.

\begin{figure}[th]
\hspace{-0.3cm}\includegraphics[width=1.04\columnwidth]{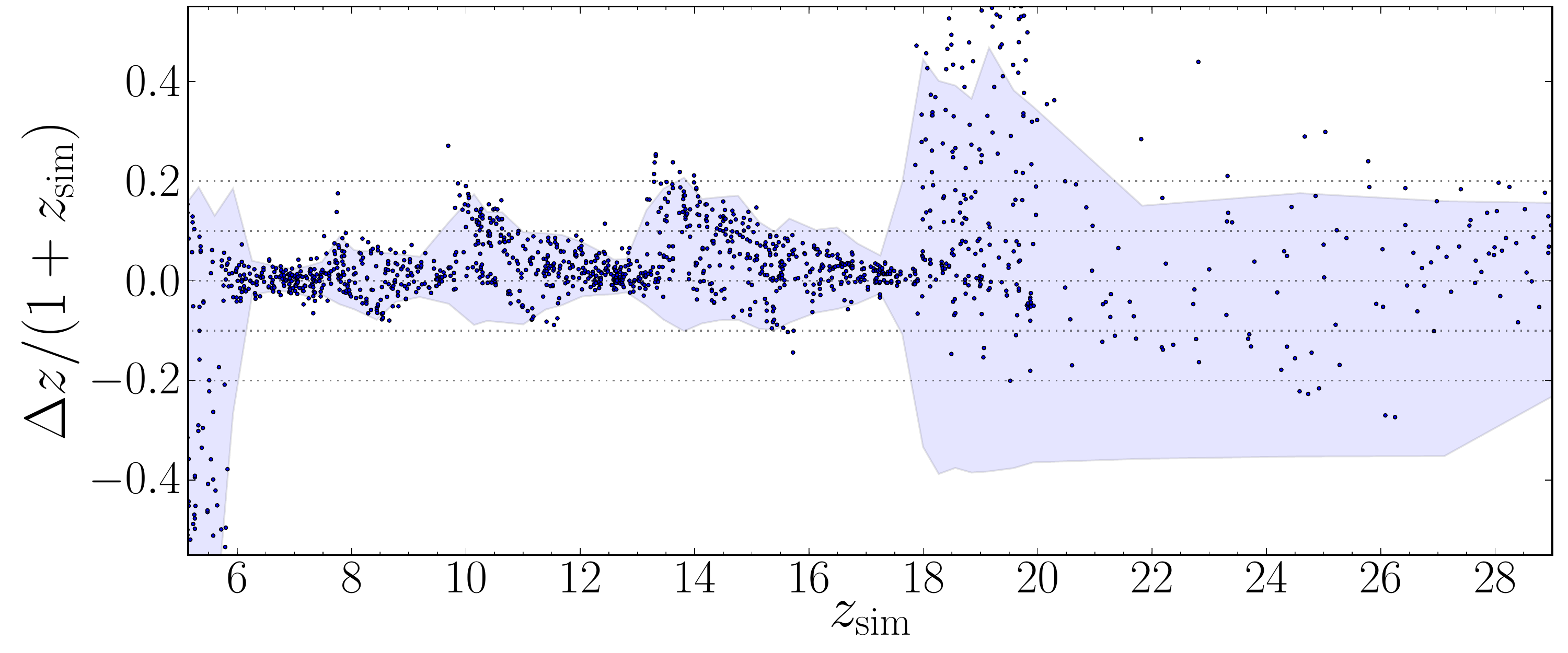}
\vspace{-0.5cm}
\caption{GRB afterglow photometric redshift accuracy of a $zYJHKLM$ filter set.
Small black dots show a mock set 
of 900 simulated afterglow spectra and their corresponding photo-$z$. 
The blue-shaded area shows the quadratic sum 
of the typical difference to the input redshift and the 1$\sigma$ statistical 
uncertainty of the photo-$z$ analysis averaged over 
30 afterglows in relative ($\eta = \Delta z/(1+z)$) terms. 
For the $7<z<17$ redshift range, the photo-$z$ can be determined to better
than 20\%. At $z > 17.5$ ($K$-dropout), the error gets larger due to
the gap above the $K$ band and the widths of the $L$ ($M$) bands;
yet, the redshift accuracy is more than sufficient for any follow-up decision.}
\label{IRTaccuracy}
\end{figure}

The inclusion of the $LM$ bands into the IRT 
requires operating temperatures of about 37 (50) K for the $M$ ($L$)
channels. This will certainly require active cooling. In addition, 
several optical elements in the optical path will have temperature
constraints, so that the thermal architecture of the instrument will
need to be designed carefully, though it will be much less stringent 
than e.g. on Herschel.

After the slew to a GRB, the XRT will provide a position with an accuracy
between 5-20\asec\, depending on the details of the XRT and the off-axis 
angle of the GRB in the XRT FOV. 
This uncertainty is too large for immediate (low-resolution)
spectroscopy, so two options are possible.

The simple and cheap option is a simultaneous multi-band imager in,
e.g., the seven bands $zYJHKLM$ \cite{grips2012}, thus covering the
redshift range $7\lsim z\lsim 30$.
Since GRB afterglow spectra are simple power laws, and at $z>3$ 
Lyman-$\alpha$ is the 
dominant spectral feature, relatively high accuracies can be reached 
even with broad-band filters (Fig.~\ref{IRTaccuracy}), 
as demonstrated in ground-based observations with GROND \cite{ksg11}.

\begin{figure*}[th]
\includegraphics[width=0.99\textwidth, clip]{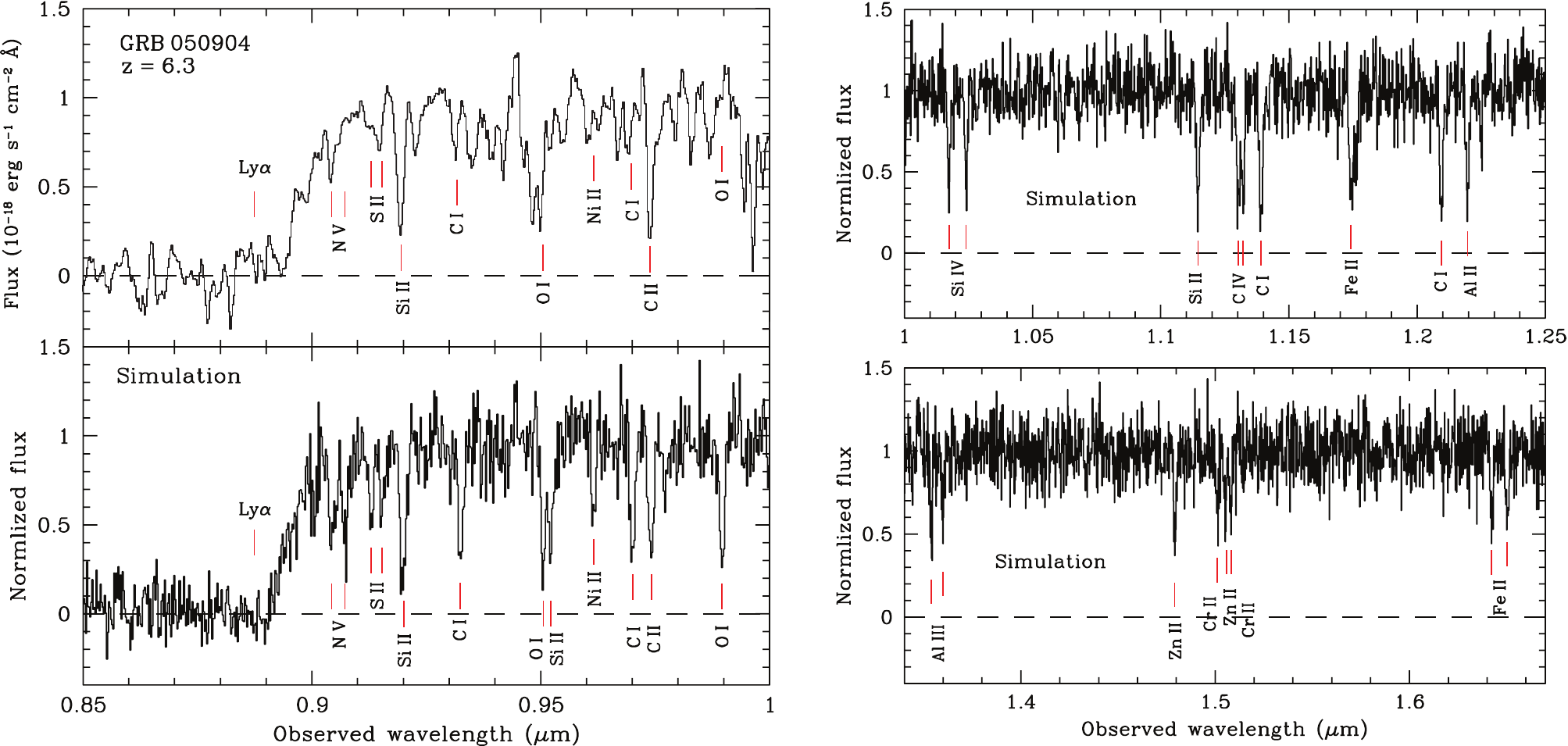}
\vspace{-0.2cm}
\caption{Spectrum of the afterglow of GRB 050904,
taken with Subaru/FOCAS 3 days after the GRB  (top left; \cite{kka06}),
and simulation of a R=1000 spectrum taken with a space-borne 1\,m telescope
for 1 hr exposure at an afterglow brightness of $J(AB)$=21 mag (lower
left panel with the same wavelength range as the observed spectrum, 
and the $J$- and $H$-band regions in the right panels).
A metallicity similar to that of GRB 050904 is assumed, with a ZnII column
of 2.5$\times$10$^{13}$ cm$^{-2}$, and ionised gas (e.g. AlIII, CIV, SiIV)
with a column of 1/10 of the neutral gas.
}
\label{grbospec}
\end{figure*}

A more sophisticated, but also more sumptuous option is a combined 
imager and spectrograph, as proposed for the dedicated GRB mission 
ORIGIN \cite{origin2012}.
Different areas of the detector are used for either
imaging in (sequentially exposed) multi-band filters,
low-resolution (R=20) spectroscopy, or high-resolution (R=1000)
integral-field spectroscopy. Switching between these modes requires
few arcmin re-pointings of the satellite, based on an accurate position derived
from initial imaging data. The power of a R=1000 NIR spectrograph on a
1\,m space telescope is demonstrated in Fig.~\ref{grbospec},
allowing unique absorption line diagnostics for $\sim$50\% of the GRBs 
up to the highest redshifts.

\section{Strawman mission concepts}

\subsection{A Distributed Approach}

As with other areas of research, the next step forward in understanding
GRBs or using them as a tool requires a substantial larger effort on
the instrumentation. The basic idea behind this distributed approach 
is our conviction that strategically
linking together future large/expensive global facilities (both ground-
and space-based) is of considerable importance to maximise the overall
scientific return, in particular at the ever growing costs with more and more
ambitious projects. In a perfect world, the different major funding
agencies could be expected to seriously weigh up the possible synergy 
in making their selections.

\paragraph{Separating the tasks:}

The GRB-finder and the two narrow-field instruments do not have to 
share the same satellite. In fact, the rapid slewing of the X-ray and
(N)IR telescopes is optimised if the flight configuration has low mass
(angular momentum). A concept study with EADS Astrium indeed showed that
a 2-satellite configuration flight would be preferable even in a LEO
(at 500--2000 km distance, not requiring precision formation flying!) 
unless the GRB-finder is very simple and  light-weight.
Thus, a strawman concept would be 
(i) one satellite with a GRB-finder, and
(ii) another satellite combining the {\em XMM} or {\em eROSITA} 
  spare with an {\em EUCLID}-sized telescope (just M1 to M3 mirrors, and
  at largely reduced optical quality and alignment requirements).
The GRB-finder with the largest impact on auxiliary science would 
be a ``super-BAT'', i.e. an octahedron-shaped satellite where all
but the Sun-facing direction contain a coded mask telescope
with 2000 cm$^2$ detector area each. 
Being placed in L2, and with no slewing required,
such a configuration would detect $\sim$1200 GRBs/yr.
The follow-up satellite would slew to each GRB and observe for 
$\approx$1--2 h minimum time. This would allow up to 15 GRBs to be
observed on a single day (occuring once or twice per year), but
on average could leave about half the observing time of the 
X-ray/(N)IR telescopes to other science areas. Data could be 
sent to the GRB-finder (or other geostationary satellites) 
from where it would be much easier to rapidly downlink
to Earth due to the fixed location in space.

\paragraph{Piggyback on ESA missions under consideration:}

An alternative option could be to add a GRB-finder to one of the ESA
missions already under discussion. This would provide Table 1 items (1)
and (2), possibly even (3) for a subset of GRBs.
Providing (3) and (4) would require either a dedicated mission or a
smaller follow-up mission.
We acknowledge that these are substantial modifications to the
existing concepts.

(1) Adding a GRB-finder to ATHENA+ (or other L2-selected mission) and  
prepare for reasonably rapid (2 h)
autonomous slewing capability:
The presently planned ATHENA+ Wide-Field Imager has a FOV large enough to
cover several-arcmin sized GRB error boxes, and the calorimeter would
provide unique X-ray absorption spectra for the line-of-sights to GRBs.
In such a scenario, a separate (N)IR telescope would be needed.

(2) Similarly, adding a GRB-finder to the 
Large Observatory For X-ray Timing, {\em LOFT} \cite{Feroci12a}:
{\em LOFT} is expected to detect of order 150 GRBs/yr, 
which is too few for the purpose proposed here. Since autonomous slewing is not
part of the {\em LOFT} concept, an enhancement of the GRB-finder
capabilities would imply that a separate satellite with the X-ray 
and (N)IR telescope would be required.

\subsection{All-in-One mission}

We see the following options (though other mixing and matching of these
components would also be possible), determined by the properties of the
GRB-finder
(field-of-view and localization accuracy). Most of these configurations, 
if not all, would benefit from an L2 orbit which therefore is taken
as the default option. Depending on the combination, up to three 
autonomous slews will be needed to achieve (N)IR spectroscopy of
the GRB afterglow.

\paragraph{Scintillator, single Lobster, NIR:}

Using a GBM-like detector with 4$\times$ larger effective area, and
20--24 modules will cover the full sky and return 1000 GRBs/yr with
locations in the 1\degs--4\degs\ range. Autonomously re-pointing a 
long focal length ($\sim$2\,m), narrow-field (8\degs$\times$8\degs) 
Lobster provides a 95\% detection rate of the X-ray afterglows,
and a position accurate to $<$0.5--1\amin.
This position is good enough for the (N)IR telescope to slew
and start 7-channel imaging and obtain a 1\asec\ position. 
Another slew would place the afterglow on the spectrograph.

\paragraph{All-sky Lobster, XRT, NIR:}

Using about a dozen short focal length (thus small effective area), 
large-FOV (30\degs$\times$30\degs) Lobster modules would detect
about 1000 GRBs/yr with locations accurate to few arcmin.
Autonomously re-point a single {\em eROSITA}-like X-ray telescope to get 
a 99\% X-ray afterglow detection rate, and localizations accurate
to $\lsim$30\asec, accurate enough for (N)IR 7-channel imaging and/or
grism spectroscopy. Possibly, a longer focal length Lobster could
replace the {\em eROSITA}-like X-ray telescope.

\paragraph{Coded mask, XRT, NIR:}

Using eight Swift/BAT-like coded mask systems in octahedron orientation,
and lowering the low-energy threshold from 20 to $\sim$10 keV would
provide 1000 GRBs/yr with locations accurate to few arcmin.
Autonomously re-point a single {\em eROSITA}-like X-ray telescope to get 
a 99\% X-ray afterglow detection rate, and localizations accurate
to 30\asec, accurate enough for (N)IR 7-channel imaging and/or
grism spectroscopy.

\paragraph{Compton, XRT, NIR:}

Two systems of half a cubic-meter Compton telescopes (e.g. \cite{grips2012}),
oriented in opposite directions, will detect about 1300 GRBs/yr, out
of which about 900 will have localisations $<$1\degs. Autonomously
re-point of seven {\em eROSITA}-like X-ray telescopes, oriented to fill
a 3\degs\ diameter FOV, will provide a 99\% X-ray afterglow detection rate
The  30\asec\ localizations are accurate enough for (N)IR 7-channel 
imaging and/or grism spectroscopy.


\clearpage

\begin{thebibliography}{}
\setlength{\itemsep}{-0.1ex plus0.1ex}
\setstretch{0.8}                

\bibitem[1]{LaR2000} Lamb D.Q., Reichart D.E., 2000, ApJ 536, L1

\bibitem[2]{nab07} Naoz S., Bromberg O., 2007, MN 380, 757

\bibitem[3]{hjo03} Hjorth J. et al. 2003, Nat 423, 847

\bibitem[4]{bbm13} Bartos I., Brady P., Marka S. 2013, Class. Quant. Grav. 30, 
123001

\bibitem[5]{slgrbs} NASA, http://ecuip.lib.uchicago.edu/multiwave\-length-astronomy/gamma-ray/science/07.html

\bibitem[6]{predehl2010} Predehl P., Andritschke R., B\"ohringer H. et al. 2010,
SPIE 7732E, 23

\bibitem[7]{kss12} Khabibullin I., Sazonov S., Sunyaev R. 2012, MN 426, 1819

\bibitem[8]{Amati2013} Amati L., Del Monte E., D'Elia V. et al. 2013, 
Nucl. Phys. B (in press; arXiv:1302.5276)

\bibitem[9]{gao12} Gendreau K.C., Arzoumanian Z., Okajima T. 2012,
SPIE 8443, 844313

\bibitem[10]{pbb13} Park I.H., Brandt S., Budtz-Jorgensen C. et al. 2013,
NJP 15, 023031

\bibitem[11]{Siellez2013} Siellez K., Bo\"er M., Gendre B. 2013, 
MN (subm.)

\bibitem[12]{Aasi2013} Aasi J., Abadie J., Abbott B.P. et al., 2013, 
arXiv:1304.0670

\bibitem[13]{Sathy12} Sathyaprakash B. et al 2012, Class. Quant. Grav. 29, 
124013

\bibitem[14]{aba10} Abadie J., et al. 2010, Class. Quant. Grav. 27, 173001

\bibitem[15]{Abbasi2012} Abbasi R., Abdou Y., Abu-Zayyad T et al. 2012,
 Nat. 484, 351

\bibitem[16]{Whitehorn2013} Whitehorn N., 2013, talk at IceCube Part. 
Astrophys. Symp., 2013 May 15, Madison

\bibitem[17]{HBW12} H\"ummer S., Baerwald P., Winter W. 2012, 
  PRL 108, 231101

\bibitem[18]{Bouwens2012} Bouwens R.J., Illingworth G.D., Oesch P.A. et al. 
2012, ApJ 752, L5

\bibitem[19]{Tanvir2012} Tanvir N., Levan A.J., Fruchter A.S. et al. 2012,
ApJ 754, 46

\bibitem[20]{john2008} Johnson J.L., Greif T., Bromm V., 2008, MN 388, 26

\bibitem[21]{rpd08} Rossi E., Perna R., Daigne F 2008, MN 390, 675

\bibitem[22]{souz12} de Souza R.S., Krone-Martins A., Ishida E.E.O., 
Ciardi B., 2012, A\&A 545, A9

\bibitem[23]{zrk13} Zhu S., Racusin J., Kocevski D. et al. 2013, GCN \#14471

\bibitem[24]{gbc13} Gilmore R.C., Bouvier A., Connaughton V. et al. 2013,
  Exp. Astron. 35, 413

\bibitem[25]{Ghirlanda13} Ghirlanda G. et al. 2013, MN (subm.)

\bibitem[26]{clf11} Cucchiara A., 
et al. 2011, ApJ 736, C7

\bibitem[27]{teg97} Tegmark M., 
et al. 1997, ApJ 474, 1

\bibitem[28]{ybh04} Yoshida N., Bromm V., Hernquist L., 2004, ApJ 605, 579

\bibitem[29]{kyk13} Kamada A., 
et al. 2013,
  JCAP 03, 008

\bibitem[30]{bho01} Barkana R., Haiman Z., Ostriker J.P., 2001, ApJ 558, 482

\bibitem[31]{mes05} Mesinger A., Perna R., Haiman Z. 2005, ApJ 623, 1

\bibitem[32]{smf13} de Souza R.S., Mesinger A., Ferrara A. et al. 2013, 
MN (subm., arXiv:1303.5060)

\bibitem[33]{Maio12} Maio U., 
et al. 2012, MN 426, 2078

\bibitem[34]{byh09} Bromm V., Yoshida N., Hernquist L., McKee C.F.,
  2009, Nat. 459, 49

\bibitem[35]{bl04} Bromm V., Larson R.B. 2004, ARA\&A 42, 79

\bibitem[36]{glo05} Glover S. 2005, SSRv 117, 445

\bibitem[37]{CiF05} Ciardi B., Ferrara A. 2005,  SSRv 116, 625

\bibitem[38]{hob06} Hopkins A.M., Beacom J.F. 2006, ApJ 651, 142

\bibitem[39]{Madau1999} Madau P., Haardt F., Rees M.J., ApJ 514, 648

\bibitem[40]{Kistler2013} Kistler M.D., Y\"uksel H., Hopkins A.M. 2013, 
arXiv:1305:1630

\bibitem[41]{abn02} Abel T., Bryan G.L., Norman M.L. 2002, Sci. 295, 93

\bibitem[42]{bcl02} Bromm V., Coppi P.S., Larson R.B. 2002, ApJ 564, 23

\bibitem[43]{yoh06} Yoshida N., Omukai K., Hernquist L., Abel T. 2006, 
ApJ 652, 6

\bibitem[44]{osn07} O'Shea B.W., Norman M.L. 2007, ApJ 654, 66

\bibitem[45]{cgs11a} Clark P.C., Glover S., Simon C.O. et al. 2011a, 
Sci. 331, 1040

\bibitem[46]{gbc12} Greif T., Bromm V., Clark P.C. et al. 2012, MN 424, 399

\bibitem[47]{sgk13} Stacy A., Greif T., Klessen R.S. et al. 2013, MN 431, 1470

\bibitem[48]{tao09} Turk M., Abel T., O'Shea B 2009, Sci 325, 601

\bibitem[49]{sgb10} Stacy A., Greif T., Bromm V. 2010, MN 403, 45

\bibitem[50]{cgs11b} Clark P.C., Glover S., Simon C.O. et al. 2011b, 
ApJ 727, 110

\bibitem[51]{sgs11} Smith R.J., Glover S., Simon C.O. et al. 2011, MN 414, 3633

\bibitem[52]{BeC05} Beers $\!$T.C., Christlieb N., 2005, $\!$ARA\&A 43, 531

\bibitem[53]{ssf07} Salvadori S., Schneider R., Ferrara A. 2007, MN 381, 647 

\bibitem[54]{Tum07} Tumlinson J., 2007, ApJ 664, 63

\bibitem[55]{tun07} Tominaga N. Umeda H., Nomoto K., 2007, ApJ 660, 516

\bibitem[56]{iut09} Izutani N., Umeda H., Tominaga N. 2009, ApJ 692, 1517

\bibitem[57]{HeW10} Heger A., Woosley S.E. 2010 ApJ, 724, 341

\bibitem[58]{jab10} Joggerst C.C. 
et al. 2010, ApJ 709, 11

\bibitem[59]{sfs03} Schneider R., Ferrara A., Salvaterra R. et al. 2003, 
Nat. 422, 869

\bibitem[60]{sol12} Schneider R., Omukai K., Limongi M. 2012, MN 423, L60

\bibitem[61]{jgb09} Johnson J.L., Greif T.H., Bromm V. et al. 2009, 
ASP Conf. 419, p. 335

\bibitem[62]{bl06} Bromm V., Loeb A., 2006, ApJ 642, 382

\bibitem[63]{Campisi2011} Campisi M.A., Maio U., Salvaterra R., 
  Ciardi B., 2011, MN 416, 2760

\bibitem[64]{MeR10} Meszaros P., Rees M.J., 2010, ApJ 715, 967

\bibitem[65]{KoB10} Kommissarov S.S., Barkov M.V., 2010, MN 402, L25

\bibitem[66]{SuI11} Suwa Y., Ioka K., 2011, ApJ 726, 107

\bibitem[67]{gkf09} Greiner J., Kr\"uhler T., Fynbo J.P.U, 
et al. 2009, ApJ 693, 1610

\bibitem[68]{2009Natur.461.1254T} Tanvir N.R., Fox D.B., Levan A.J. et al.\
2009, Nat. 461, 1254

\bibitem[69]{2009Natur.461.1258S} Salvaterra R., Della Valle M., Campana S. et
al. 2009, Nat. 461, 1258

\bibitem[70]{fwh99} Fryer C., Woosley S.E., Hartmann D.H. 1999, ApJ 526, 152

\bibitem[71]{fps08} Fynbo J.P.U., 
et al.   2008, ApJ 683, 321

\bibitem[72]{KYB09} Kistler M.D., 
et al. 2009, ApJ 705, L104

\bibitem[73]{egk12} Elliott J., Greiner J., Khochfar S., 
et al. 2012, A\&A 539, A113

\bibitem[74]{Daigne2006} Daigne F., Rossi E.M., Mochkovitch R. 2006, 
MN 372, 1034

\bibitem[75]{wap10} Wanderman D., Piran T. 2010, MN 406, 1944

\bibitem[76]{isf11} Ishida E.E.O., de Souza R., Ferrara A. 2011, 
MN 418, 500

\bibitem[77]{gkk11} 
Greiner J.,  
et al. 2011, A\&A 526, A30

\bibitem[78]{scv12} Salvaterra R., Campana S., Vergani S.D. 
  et al. 2012, ApJ 749, 68:

\bibitem[79]{wbg12} Wang F.Y., Bromm V., Greif T. et al. 2012, ApJ 760, 27 

\bibitem[80]{sfs04} Schneider R., Ferrara A., Salvaterra R. et al. 2004,
 MN 351, 1379

\bibitem[81]{vel04} Vreeswijk P.M., 
et al. 2004, 
A\&A 419, 927 

\bibitem[82]{Black98} Black J.H., 1998, Faraday Discuss. 109, 257

\bibitem[83]{mao98} Mao S., Mo H.J., 1998, A\&A 339, L1

\bibitem[84]{ChF06} Choudhury T.,  Ferrara A. 2006, 
MN 371, 55

\bibitem[85]{ABS06} Alvarez M.A. et al., 2006,
ApJ 639, 621

\bibitem[86]{mcf11} Mitra S., Choudhury T., Ferrara A. 2011, 
MN 419, 1480

\bibitem[87]{SSR04} Simcoe R.A. et al., 2004,
ApJ 606, 92

\bibitem[88]{tfs07} Tornatore L., Ferrara A.,  Schneider R. 2007, 
MN 382, 945

\bibitem[89]{smc13} Salvaterra R., Maio U., Ciardi B., et al. 2013, MN
429, 2718

\bibitem[90]{kka06} Kawai N., 
 et al. 2006, 
  Nat. 440, 184

\bibitem[91]{Ellis2013} Ellis R.S., 
et al. 2013, ApJ 763, L7

\bibitem[92]{Basa2012} Basa S., 
et al. 2012, A\&A 542, A103

\bibitem[93]{2009ApJ...691..182S} Savaglio S., Glazebrook K., Le Borgne D.
2009, ApJ, 691, 182

\bibitem[94]{bouwens2011} Bouwens R.J., 
et al. 2011, ApJ 737, 90

\bibitem[95]{npc12} Noterdaeme P., Petitjean P., Carithers W.C. et al. 2012, 
  A\&A 547, L1

\bibitem[96]{2006AA...460L..13J} Jakobsson P., Fynbo J.P.U., Ledoux C., et
al. 2006, A\&A 460, L13

\bibitem[97]{2009ApJS..185..526F} Fynbo J.P.U., Jakobsson P., Prochaska J.X.,
et al. 2009, ApJS 185, 526

\bibitem[98]{2012ApJ...752...62J} Jakobsson P., Hjorth J., Malesani D., et al.
2012, ApJ 752, 62

\bibitem[99]{gff08} Gallerani S., Ferrara A., Fan X., Choudhury T. 2008, 
MN 386, 359

\bibitem[100]{mes10} Mesinger A., 2010, MN 407, 1328

\bibitem[101]{mcg10} McGeer I.D., Mesinger A., Fan X. 2011, MN 415, 3237

\bibitem[102]{ciardi12} Ciardi B. et al. 2012, MN 423, 558

\bibitem[103]{BaL04} Barkana R., Loeb A., 2004, ApJ 601, 64

\bibitem[104]{McQ08} McQuinn M., 
et al. 2008, MN 388, 1101

\bibitem[105]{Mes04} Mesinger A. et al. 2004, ApJ 613, 23

\bibitem[106]{Tot06} Totani T., Kawai N., Kosugi G. et al. 2006, PASJ 58, 485

\bibitem[107]{MeF08} Mesinger A., Furlanetto S.R. 2008, MN 385, 1348

\bibitem[108]{Ino10} Inoue S., Salvaterra R., Choudhury T.R. et al. 2010, 
MN 404, 1938

\bibitem[109]{Camp10} Campana S., Th\"one C.C., de Ugarte Postigo A. 
et al. 2010, MN 402, 2429

\bibitem[110]{Camp12} Campana S., Salvaterra R., Melandri A. 
et al. 2012, MN 421, 1697

\bibitem[111]{WaJ12} Watson D., Jakobsson P. 2012, ApJ 754, 89


\bibitem[112]{Behar11} Behar E., 
et al. 2011, 
ApJ 734, 26

\bibitem[113]{Star13} Starling R.L.C., Willingale R., Tanvir N.R. 
et al. 2013, MN (acc.; arXiv:1303.0844)

\bibitem[114]{Nic05} Nicastro F., 
et al. 2005, ApJ 629, 700

\bibitem[115]{Camp11} Campana S., Salvaterra R., Tagliaferri G.  
et al. 2011, MN 410, 1611

\bibitem[116]{becker08} Becker J.K., 2008, Phys. Rep. 458, 172

\bibitem[117]{WaB97} Waxman E., Bahcall J. 1997, PRL 78, 2292

\bibitem[118]{bloom09} Bloom J.S. et al. 2009, Astro2010 White paper, 
arXiv:0902.1527

\bibitem[119]{phinney09} Phinney E.S. 2009, Astro2010 White paper, 
arXiv:0903.0098

\bibitem[120]{Coward2012} Coward D.M., 
et al., 2012, 
MN 425, 2668

\bibitem[121]{kli11} Klimenko S. et al., 2011, PhRvD 83, 102001

\bibitem[122]{dalal06} Dalal N. et al., 2006, PhRvD 74, 063006

\bibitem[123]{Sathy10} Sathyaprakash B. et al 2010, Class. Quant. Grav. 27, 
215006

\bibitem[124]{LiP98} Li L.-X., Paczy{\'n}ski B. 1998, ApJ 507, L59

\bibitem[125]{Dessart09} Dessart L., 
et al. 2009, 
ApJ 690,  1681

\bibitem[126]{pnr13} Piran T., Nakar E., Rosswog S. 2013, MN 430, 2121
ack98

\bibitem[127]{SaSchu2009} Sathyaprakash B.S., Schutz B.F., 2009, Liv. Rev. Rel. 12, No. 2

\bibitem[128]{cut02} Cutler C., Thorne K.S., 2002, in: 16th GRG Conference, 
World Sci. Publ., p. 72

\bibitem[129]{Granot03} Granot J.  2003, ApJ 596, L17

\bibitem[130]{Toma09} Toma K., 
et al. 2009,  ApJ 698, 1042

\bibitem[131]{gk03} Granot J., K\"onigl A., 2003, ApJ 594, L83

\bibitem[132]{lyu03} Lyutikov M., Pariev V.I., Blandford R.D. 2003, ApJ 597, 998

\bibitem[133]{waxman03} Waxman E., 2003, Nat. 423, 388

\bibitem[134]{Belo11} Beloborodov A.M. 2011, ApJ 737, 68

\bibitem[135]{kalemci07}
{Kalemci} E., {Boggs} S.~E., {Kouveliotou} C., 
et al. 2007, ApJ Suppl. 169, 75 

\bibitem[136]{McGlynn07} McGlynn S., Clark D.J., Dean A.J.  et al. 2007, A\&A 
466, 895 

\bibitem[137]{mcglynn09} McGlynn S. et al., 2009, 
A\&A 499, 465

\bibitem[138]{gotz09}
{G\"otz} D., {Laurent} P., {Lebrun} F. et al.
2009, ApJ 695, L208

\bibitem[139]{yonetoku11}
{Yonetoku} D. et al., 2011, ApJ 743, L30

\bibitem[140]{yonetoku12}
{Yonetoku} D. et al., 2012, ApJ 758, L1

\bibitem[141]{gotz13}
{G\"otz} D., Covino S. Fernandez-Soto A. et al.
2013, MN (in press, arXiv:1302.4186)

\bibitem[142]{fan07}
Fan Y.-Z., Wei D., Xu D. $\!$2007, MN 376, 1857

\bibitem[143]{laurent11a}
{Laurent} P., {G{\"o}tz} D., {Bin{\'e}truy} P. et al.
2011, Phys. Rev. D83, 12, 121301

\bibitem[144]{toma12}
Toma K. et al., 2012, Phys. Rev. D109, 24

\bibitem[145]{GalYam12} Gal-Yam A. 2012, Sci. 337, 927

\bibitem[146]{bloom11} Bloom J.S., Giannios D., Metzger B.D. et al. 201, 
Sci. 333, 203

\bibitem[147]{kom12} Komossa S. 2012, EPJ Web Conf. 39, id. 02001

\bibitem[148]{mlb09} Meegan C., 
et al. 2009, ApJ 709, 791

\bibitem[149]{hpm11} Hudec R., Pina L., Marsikova V., 
et al. 2011, AIP Conf. Proc. 1358, p. 423

\bibitem[150]{tichy11} Tichy V. et al. 2011, NIM Phys. Res. A. 633, p. S169

\bibitem[151]{bfp12} Burrows D.N., Fox D., Palmer D. et al. 2012, 
Mem. S.A. It. 21, p. 59

\bibitem[152]{Gehr12} Gehrels N., Barthelmy S.D., Cannizzo J.K., 2012, 
Proc. IAU Symp. 285, p. 41

\bibitem[153]{grips2012} Greiner J., Mannheim K., Aharonian F. et al. 2012,
Exp. Astr. 34, 551

\bibitem[154]{ksg11}
Kr\"uhler T., Schady P., Greiner J. et al. 2011, A\&A 526, A153

\bibitem[155]{origin2012} den Herder J.-W., Piro L., Ohashi T. et al. 2012,
Exp. Astr. 34, 519


\bibitem[156]{Feroci12a} Feroci M., Stella L., van der Klis M. et al. 2012,
 Exp. Astr. 34, 415

\end{thebibliography}
\end{document}